\documentclass[a4paper,twocolumn,accepted=2026-04-14,10pt]{quantumarticle}
\pdfoutput=1
\usepackage[colorlinks,linkcolor=blue,citecolor=blue]{hyperref}
\usepackage{xcolor}
\usepackage{amsmath,amssymb}
\usepackage{commath}
\usepackage{bm}
\usepackage{braket}
\usepackage{mathtools}
\usepackage{physics}
\usepackage{cleveref}
\usepackage{graphicx}
\usepackage{epstopdf}
\usepackage{bbm}
\usepackage{CJKutf8}

\usepackage[backend=bibtex,sorting=none,isbn=false,doi=false,url=false,citestyle=numeric-comp,maxbibnames=4]{biblatex}
\DeclareFieldFormat[article]{title}{
  \iffieldundef{url}
    {#1}
    {\href{\thefield{url}}{#1}}}

\usepackage{tikz}
\usetikzlibrary{decorations.pathreplacing,calligraphy,decorations.markings}

\definecolor{tensor}{rgb}{0.5,0.8,0.5}
\definecolor{isometry}{rgb}{0.8,0.8,1}
\definecolor{unitary}{rgb}{0.8,0.5,.5}
\definecolor{gate}{rgb}{1.0,1.0,1.0}

\newcommand{\ATensor}[2]{
	\begin{scope}[shift={(#1)}]
		\draw (-1,0) -- (1,0);
		\draw (0,1) -- (0,0);
		\filldraw[fill=tensor] (-1/2,-1/2) -- (-1/2,1/2) -- (1/2,1/2) -- (1/2,-1/2) -- (-1/2,-1/2);
		\draw (0,0) node {\scriptsize #2};
	\end{scope}
}
\newcommand{\ADaggTensor}[2]{
	\begin{scope}[shift={(#1)}]
		\draw (-1,0) -- (1,0);
		\draw (0,-1) -- (0,0);
		\filldraw[fill=tensor,shift={(0,0)}] (-1/2,-1/2) -- (-1/2,1/2) -- (1/2,1/2) -- (1/2,-1/2) -- (-1/2,-1/2);
		\draw (0,0) node {\scriptsize #2};
	\end{scope}
}
\newcommand{\PTensor}[1]{
	\begin{scope}[shift={(#1)}]
		\draw (-1.5,0) -- (1.5,0);
		\draw (-.6, 0) -- (-.6, 1);
		\draw ( .6, 0) -- ( .6, 1);
		\filldraw[fill=tensor] (-1,-1/2) -- (-1,1/2) -- (1,1/2) -- (1,-1/2) -- (-1,-1/2);
		\draw (0,0) node {\scriptsize $P_{\rm RG}$};
	\end{scope}
}
\newcommand{\BTensor}[2]{
	\begin{scope}[shift={(#1)}]
		\draw (-1,0) -- (1,0);
		\foreach \x in {0,1,...,3}{
			\draw[shift={(-.3+0.2*\x,0)}] (0,1) -- (0,0);
		}
		\filldraw[fill=tensor] (-1/2,-1/2) -- (-1/2,1/2) -- (1/2,1/2) -- (1/2,-1/2) -- (-1/2,-1/2);
		\draw (0,0) node {\scriptsize #2};
	\end{scope}
}
\newcommand{\WTensor}[2]{
	\begin{scope}[shift={(#1)}]
		\draw (-1.5,0) -- (1.5,0);
		\foreach \x in {0,1,...,3}{
			\draw[shift={(-.6+0.4*\x,0)}] (0,1) -- (0,0);
		}
		\filldraw[fill=tensor] (-1,-1/2) -- (-1,1/2) -- (1,1/2) -- (1,-1/2) -- (-1,-1/2);
		\draw (0,0) node {\scriptsize #2};
	\end{scope}
}

\newcommand{\isometry}[2]{
	\begin{scope}[shift={(#1)}]
		\foreach \x in {0,1,...,3}{
			\draw[shift={(-.6+0.4*\x,0)}] (0,1) -- (0,0);
		}
		\foreach \x in {0,1,...,1}{
			\draw[shift={(-.6+1.2*\x,0)}] (0,-1) -- (0,0);
		}
		\filldraw[fill=isometry] (-1,-1/2) -- (-1,1/2) -- (1,1/2) -- (1,-1/2) -- (-1,-1/2);
		\draw (0,0) node {\scriptsize #2};
	\end{scope}
}
\newcommand{\unitary}[1]{
	\begin{scope}[shift={(#1)}]
		\def\xscale{2}
		\foreach \x in {0,1,...,3}{
			\draw[shift={(-.6*\xscale+0.4*\x*\xscale,0)}] (0,1) -- (0,-1);
		}
		\filldraw[fill=unitary] (-\xscale,-1/2) -- (-\xscale,1/2) -- (\xscale,1/2) -- (\xscale,-1/2) -- (-\xscale,-1/2);
		\draw (0,0) node {\scriptsize$U_{\rm RG}$};
		\draw (-.2*\xscale,-1.3) node {\scriptsize$0$};
		\draw (+.2*\xscale,-1.3) node {\scriptsize$0$};
	\end{scope}
}

\newcommand{\PDaggTensor}[2]{
	\begin{scope}[shift={(#1)}]
		\draw (-1.5,0) -- (1.5,0);
		\draw (-.6, 0) -- (-.6, -1);
		\draw ( .6, 0) -- ( .6, -1);
		\filldraw[fill=tensor] (-1,-1/2) -- (-1,1/2) -- (1,1/2) -- (1,-1/2) -- (-1,-1/2);
		\draw (0,0) node {\scriptsize $#2$};
	\end{scope}
}

\definecolor{purple}{rgb}{.6,.1,.6}
\definecolor{darkgreen}{rgb}{.1,.6,.1}

\begin{document}
\crefname{equation}{Eq.}{Eqs.}
\crefname{figure}{Fig.}{Fig.}

\graphicspath{{./figs/}}

\title{State preparation with parallel-sequential circuits}
\author{Zhi-Yuan Wei \begin{CJK}{UTF8}{gbsn}(魏志远)\end{CJK}}
\email{zywei@umd.edu}
\affiliation{
Max-Planck-Institut f{\"{u}}r Quantenoptik, Hans-Kopfermann-Str. 1, 85748 Garching, Germany
}

\affiliation{
Munich Center for Quantum Science and Technology (MCQST), Schellingstr. 4, 80799 M{\"{u}}nchen, Germany
}

\affiliation{
Joint Quantum Institute and Joint Center for Quantum Information and Computer Science, NIST/University of Maryland, College Park, Maryland 20742, USA
}

\author{Daniel Malz}
\email{daniel.malz@unibas.ch}
\thanks{Current affiliation: University of Basel, Klingelbergstrasse 82, CH-4056}
\affiliation{Department of Mathematical Sciences, University of Copenhagen, 2100 Copenhagen, Denmark
}

% \date{June 13, 2024}
\begin{abstract}
We introduce parallel-sequential (PS) circuits, a family of quantum circuit layouts that interpolate between brickwall and sequential circuits, which introduces control parameters governing a trade-off between the amount of entanglement and the maximum correlation range they can express. We provide numerical evidence that PS circuits can efficiently prepare many-body ground states in one dimension. On noisy devices, characterized through both idling errors and two-qubit gate errors, we show that in a wide parameter regime, PS circuits outperform brickwall, sequential, and the log-depth circuits from [Malz, Styliaris, Wei, Cirac, PRL 132, 040404 (2024)]. Additionally, we demonstrate that properly chosen noisy random PS circuits suppress error proliferation and, when employed as a variational ansatz, exhibit superior trainability.

\end{abstract}

\maketitle

\section{Introduction}Quantum circuits are a unified framework to describe operations on quantum processors~\cite{Nielsen2000}, but can also serve as a model of many-body dynamics. In this vein, different circuit layouts (arrangement of gates) correspond to distinct processes and can be used to define different variational families of states.
For example, brickwall circuits naturally model evolution with spatial locality~\cite{Vidal2004} and thus efficiently prepare states obtained in a quench from a product state. In contrast, sequential circuits prepare states with constant entanglement entropy, which in one dimension includes all matrix product states (MPS)~\cite{Schon2005} and in higher dimensions a rich family of states including topologically ordered ones~\cite{Banuls2008,Zaletel2020,zypp,yj_circ,chen2024sequential}. The multi-scale entanglement renormalization ansatz (MERA)~\cite{Vidal2006} can model critical states and can be interpreted as arising from holographic dynamics.

The advantage of using a variational family that is equivalent to a quantum circuit applied to a product state is that one can directly prepare states in that family on quantum devices. This approach is thus widely used to prepare quantum states represented as tensor network states by converting (or compiling) them into a circuit~\cite{smith2022crossing,Ran2020a,lin2021real,haghshenas2022variational,rudolph2023decomposition}, but also in variational quantum algorithms that employ a quantum computer to optimize with a family defined through a circuit layout (usually brickwall circuits)~\cite{cerezo2021variational,bharti2022noisy,tilly2022variational}.

When preparing states on noisy devices, the circuit that achieves the highest fidelity in the noiseless setting (has the lowest \emph{approximation error}) may no longer be optimal. \emph{Idling errors} accumulate with circuit depth and \emph{gate errors} with the number of gates. The optimal circuit must balance these three sources of error. This depends on the architecture considered. For concreteness, we mainly consider one-dimensional devices with spatially local gates, but in \cref{apd_hdim} we also introduce a higher-dimensional generalization. In the one-dimensional setting, sequential circuits and MERA have a depth (in terms of local gates) linear in system size $N$ and are therefore very susceptible to idling error. Brickwall circuits are the densest possible arrangement of spatially local gates and thus achieve the lowest depth for any task, but may use too many gates. If the goal is to prepare gapped ground states, which have constant (in system size) entanglement entropy~\cite{hastings2007area,eisert2010colloquium}, as well as exponentially decaying correlations~\cite{PhysRevLett.93.140402}, one might heuristically expect that the optimal circuit has logarithmic depth, but a constant number of gates crossing any particular bipartition of the chain, which is fulfilled by none of the aforementioned layouts. While the \textit{RG circuits} in Ref.~\cite{malz2023preparation} can be used
to prepare gapped ground states in terms of MPS with such an asymptotical scaling, they incur significant overhead when compiled into two-qubit gates [cf.~\cref{ps_rg_compare}].

In this work, we introduce \textit{parallel-sequential} (PS) circuits, designed to manage this trade-off optimally. PS circuits interpolate between brickwall and sequential circuits, with tunable degree of entanglement and correlation range. In contrast to sequential circuits, PS circuits are shallow, which suppresses idling errors, and compared to brickwall circuits they use fewer gates, which mitigates gate errors. Focusing on one dimension (see \cref{apd_hdim} for higher-dimensional generalizations), we find that PS circuits are sufficiently expressive, more noise robust than other circuits, and are also easier to find variationally. Specifically, we first show numerically that in the absence of noise, single-layer PS circuits can represent short-range correlated MPS of bond dimension $D = 2$ with a small infidelity $\varepsilon$ using a circuit depth $T \sim \log(N/\varepsilon)$ that is substantially shallower than the state-of-the-art RG circuits~\cite{malz2023preparation}. Then we provide evidence that the PS circuits outperform the brickwall and sequential circuits for finding ground state of the XY model in noisy devices. To further support our findings, we analyse random PS circuits and show that those layouts with a constant number of layers have better trainability and less error proliferation compared to brickwall circuits of the same depth and sequential circuits.

\begin{figure}[h!]
	\centering
    	\includegraphics[width=0.48\textwidth]{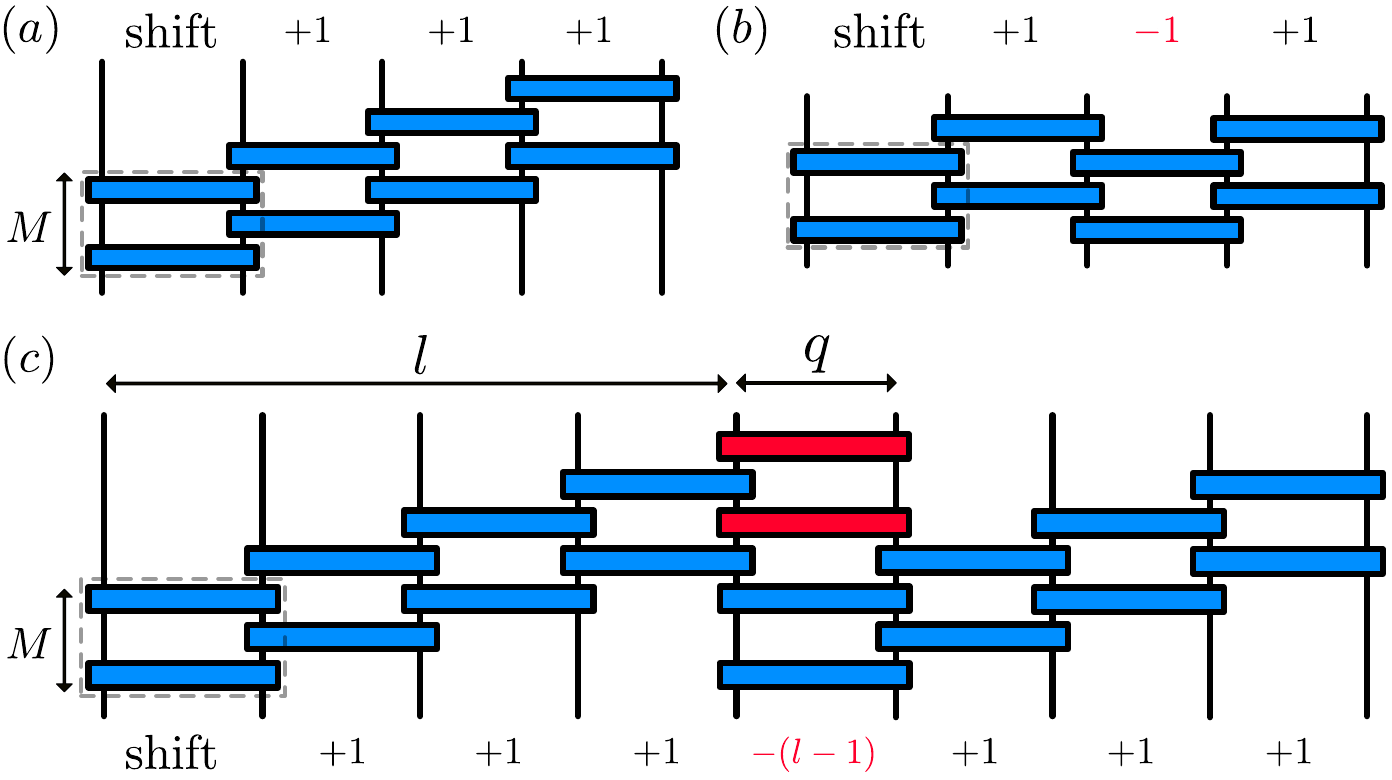}
        \caption{\textbf{(a) Sequential, (b) Brickwall, and (c) Parallel-Sequential (PS) Circuits.} The sequential and brickwall circuits both apply an identical stack of $M$ gates (dashed boxes) to each pair of neighboring qubits. The distinction lies in the shift pattern: in sequential circuits, each successive stack is shifted upward by one position (`$+1$') relative to the previous one, whereas in brickwall circuits, the stacks alternate between being shifted up (`$+1$') and down (`$-1$'). In PS circuits (panel c), the stack is shifted upward $l$ times—forming a sequential segment—followed by a single downward shift of $l-1$ steps, creating a break. To mitigate this break, we allow sequential segments to overlap, such that adjacent segments share a region of $q$ gates (highlighted in red; $q = 2$ in the figure).}
        \label{psc_illu}
\end{figure}

\section{Definition} Before we introduce 1D parallel-sequential circuits, we note that $M$-layer sequential circuits [cf.~\cref{psc_illu}(a)] and depth-$2M$ brickwall circuits [cf.~\cref{psc_illu}(b)] can both be viewed as placing the same stack of $M$ gates on each pair of neighbouring qubits, but with the difference that in sequential circuits, each subsequent stack is shifted up relative to its predecessor, whereas in brickwall circuits, they are alternately shifted up and down. In both cases, the entanglement entropy across any cut is bounded by $S \leq 2M$, but the correlations in a brickwall circuit extend only over $4M$ sites, whereas in a sequential circuit, they extend over the entire system.

In PS circuits [cf.~\cref{psc_illu}(c)], we take a mixed approach: the stack is shifted up $l$ times (generating a sequential chunk) and then once down by $l-1$ (a break). To smoothen the break, we allow the sequential chunks to extend, such that neighboring chunks overlap over a distance $q$ (red gates, $q=2$ in the figure).
This defines PS circuits with $n_C=\lceil N/l \rceil$ chunks and circuit depth $T=l+q+2M-3$. In the rest of this article, we understand brickwall and sequential circuits as opposite limits of PS circuits, where $M$-layer PS circuits with $l=2, q=1$, coincide with depth-$2M$ brickwall circuits, while those with $l=N-1, q=1$ are $M$-layer sequential circuits.

With gate parameters $\bm{\theta}$ and layout hyperparameters $N,M,l,q$, PS circuits can be expressed as a unitary $U_{\rm PS}(\bm{\theta}, N, M, l, q)$, which also defines the associated variational family
\begin{equation} \label{ps_state}
|\Psi_{\rm PS}(\bm{\theta}, N, M, l, q)\rangle = U_{\rm PS}(\bm{\theta}, N, M, l, q)|0\rangle^{\otimes N}.
\end{equation}
The entanglement entropy of $|\Psi_{\rm PS}\rangle$ across any non-overlapping cut is bounded by $S \leq 2M$, and the maximal correlation range can be tuned via $M$, $l$, and $q$ [cf.~\cref{apd_cor_fid}]. By choosing appropriate scaling of parameters, PS circuits can create states with diverse entanglement and correlation scalings. For instance, by choosing constant $M$ and $l, q \sim \log N$, PS circuits produce states with the same entanglement and correlation scaling as gapped 1D ground states.

\section{PS circuits for quantum state preparation}MPS (with open boundary conditions) of a chain of $N$ qubits and with bond dimension $D$ are defined as
\begin{equation}\label{mps_form}
|\phi_{\rm MPS}\rangle = \sum_{i_{1}, \ldots ,i_{N}=0} ^{1} A_{1}^{i_{1}} \ldots A_{N}^{i_{N}}|i_{1} \ldots i_{N}\rangle,
\end{equation}
where $\{A_n^i \}$ in the bulk are matrices of dimension $D \times D$. Typical MPS exhibit short-range correlations~\cite{garnerone2010statistical,haferkamp2021emergent,lancien2022correlation} and correspond to the ground states of local gapped Hamiltonians~\cite{Perez-Garcia2006}.
Please see a more detailed review of MPS in \cref{apd_mps_review}.

\subsection{Preparing MPS of $D=2$ with single-layer PS circuits}

\begin{figure}[h!]
\includegraphics[width=0.48\textwidth]{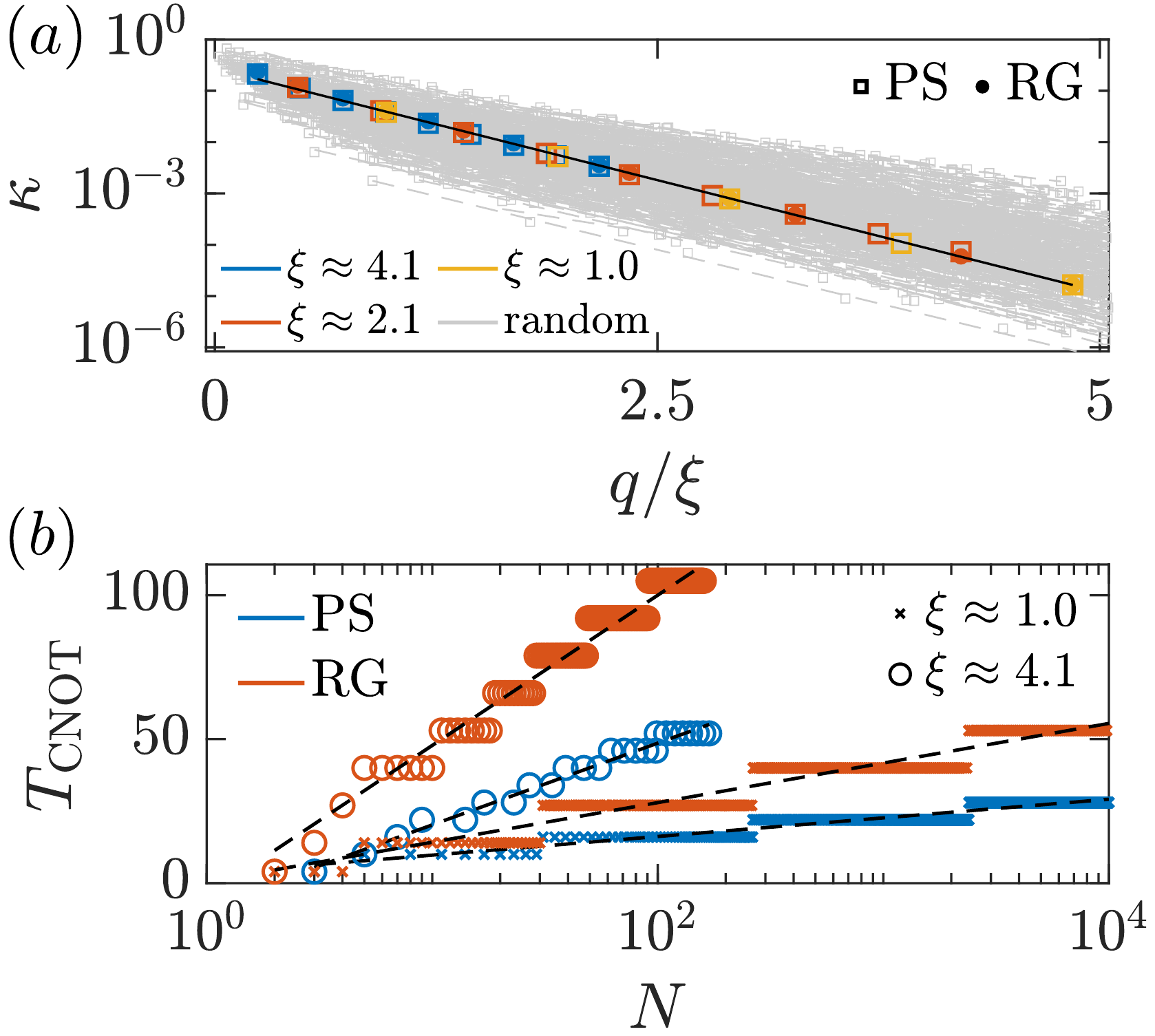}
        \caption{\textbf{Single-layer PS circuits for MPS preparation.} (a) The error density $\kappa$ as a function of $q/\xi$, for the preparation of a family of bulk-TI MPS with various correlation lengths $\xi$~\cite{Wolf2006}, as well as for 500 random bulk-TI MPS with $D = 2$, using PS circuits (hollow squares) and the RG circuit~\cite{malz2023preparation} (dots). The lines are exponential fits to the corresponding data from the same MPS. (b) Lower bound of the CNOT gate depth $T_{\rm CNOT}$ for the PS circuit and the RG circuit for preparing the MPS family of size $N$ with fidelity ${\cal F}=0.95$. The dashed lines are log fits of the data.}
        \label{exp_pow_fig}
\end{figure}

It is known that an MPS with bond dimension $D = 2$ can be prepared by a single-layer ($M = 1$) sequential circuit acting on a product state of qubits~\cite{Schon2005}, with a circuit depth that scales linearly with the system size $N$. In the following, we show that single-layer PS circuits can approximate short-range correlated MPS with $D = 2$ using exponentially smaller depth.
For simplicity, we consider MPS with identical tensors $A_i=A$, which we refer to as bulk translation-invariant (bulk-TI) MPS and denote as $| \phi_{\rm MPS}^{\textrm{bulk-TI}} \rangle$ [cf.~\cref{apd_mps_review}]. We determine the optimal approximation $|\Psi_{\rm PS} \rangle$ within the single-layer PS circuits to the target state $| \phi_{\rm MPS}^{\textrm{bulk-TI}} \rangle$ by locally optimizing the gates in the overlap region. Each overlap region introduces an error, such that the state fidelity ${\cal F} \equiv |\langle \Psi_{\rm PS} | \phi_{\rm MPS}^{\textrm{bulk-TI}} \rangle |^2$ decays exponentially with the number of chunks $n_C$ as [cf.~\cref{apd_cor_fid}]
\begin{equation} \label{}
{\cal F}(n_C, q) \approx \exp[-\kappa(q) \cdot (n_C - 1)].
\end{equation}
We numerically establish that the error density $\kappa(q)$ exhibits a generic scaling [cf.~\cref{exp_pow_fig}(a)]
\begin{equation} \label{}
\kappa(q) \sim \exp(-\gamma \cdot q / \xi).
\end{equation}
 We test this for a family of bulk-TI MPS with tunable correlation length $\xi$~\cite{Wolf2006} (the constant factor $\gamma \approx 2$ for this family), as well as random bulk-TI MPS with $D = 2$ [cf.~\cref{apd_mps_review}]. This behavior indicates that the infidelity originates from the absence of correlations in PS circuits near the location where the stack shifts down [cf.~\cref{psc_illu}(c)] [cf.~\cref{apd_cor_fid}]. Notably, the exponential decay of $\kappa(q)$ quantitatively matches the scaling observed in the RG circuit~\cite{malz2023preparation}, suggesting that the PS circuit is asymptotically optimal.

The scaling of ${\cal F}(n_C, q)$ and $\kappa(q)$ leads to a favorable circuit depth scaling, given by 
\begin{equation} \label{}
T \sim \xi \cdot \log (N / \epsilon),	
\end{equation}
for preparing $|\phi_{\rm MPS}^{\textrm{bulk-TI}}\rangle$ of size $N$ and bond dimension $D=2$ with a small infidelity $\epsilon$ using the single-layer PS circuit. From this, we estimate a lower bound on the CNOT gate depth $T_{\rm CNOT}$ after compiling the circuit [cf.~\cref{exp_pow_fig}(b)], and compare it with that from the RG circuits~\cite{malz2023preparation}. Notably, PS circuits consistently achieve a lower $T_{\rm CNOT}$ than RG circuits, as the latter involve gates acting on $1+ \lceil 2 \log _2 D \rceil$ qubits for preparing an MPS of bond dimension $D$. This results in a compilation overhead for RG circuits that increases $T_{\rm CNOT}$, which becomes more significant when preparing MPS with higher bond dimensions [cf.~\cref{ps_rg_compare}].

\subsection{Preparation of states with more entanglement}By increasing the number of layers \( M \), PS circuits (and the two limiting cases, brickwall and sequential circuits) can represent quantum states with higher entanglement. Multi-layer structures have been widely explored in the case of brickwall and sequential circuits and generally have been very effective~\cite{Bravo-Prieto2020,tilly2022variational,Ran2020a,lin2021real,rudolph2023decomposition,iqbal2022preentangling,gundlapalli2022deterministic,melnikov2023quantum,iaconis2024quantum,bohun2024scalable,gonzalez2024efficient,sano2024quantum}.
In particular, the task of preparing short-range correlated states with higher entanglement can be decomposed into a sequence of preparations of MPS with bond dimension $D = 2$~\cite{Ran2020a}. Therefore, our results for $D = 2$ MPS [cf.~\cref{exp_pow_fig}] may suggest that multi-layer PS circuits are likewise effective for representing short-range correlated states with higher entanglement. To demonstrate this more concretely, we investigate the ground-state preparation of the XY model as an illustrative example in the following. Since the XY model is gapless and exhibits power-law decaying correlations, preparing its ground state is expected to be more challenging than for gapped models. This model has been used in several studies~\cite{Ran2020a,Zhou2021} as a benchmark for ground-state preparation using multi-layer sequential circuits (a subclass of the PS circuits studied here).

The Hamiltonian of the XY model is 
\begin{equation} \label{}
H_{XY} = \sum_{i=1}^{N-1} \sigma_{i}^{x} \sigma_{i+1}^{x} + \sigma_{i}^{y} \sigma_{i+1}^{y},	
\end{equation}
where $\sigma_{i}^{x,y,z}$ are the Pauli matrices at site $i$. We quantify the state-preparation performance using the energy density above the ground state, defined as
\begin{equation} \label{}
	\nu_{\rm XY} \equiv (E - E_{\rm GS}) /N,
\end{equation}
where $E_{\rm GS}$ is the exact ground state energy. We employ a DMRG-like circuit optimization algorithm~\cite{evenbly2009algorithms} to find the optimal PS state $|\Psi_{\rm PS}\rangle$ that minimizes the energy $E = \langle \Psi_{\rm PS} | H_{XY} | \Psi_{\rm PS} \rangle$.

\begin{figure}[h!]
	\centering
	\includegraphics[width=0.48\textwidth]{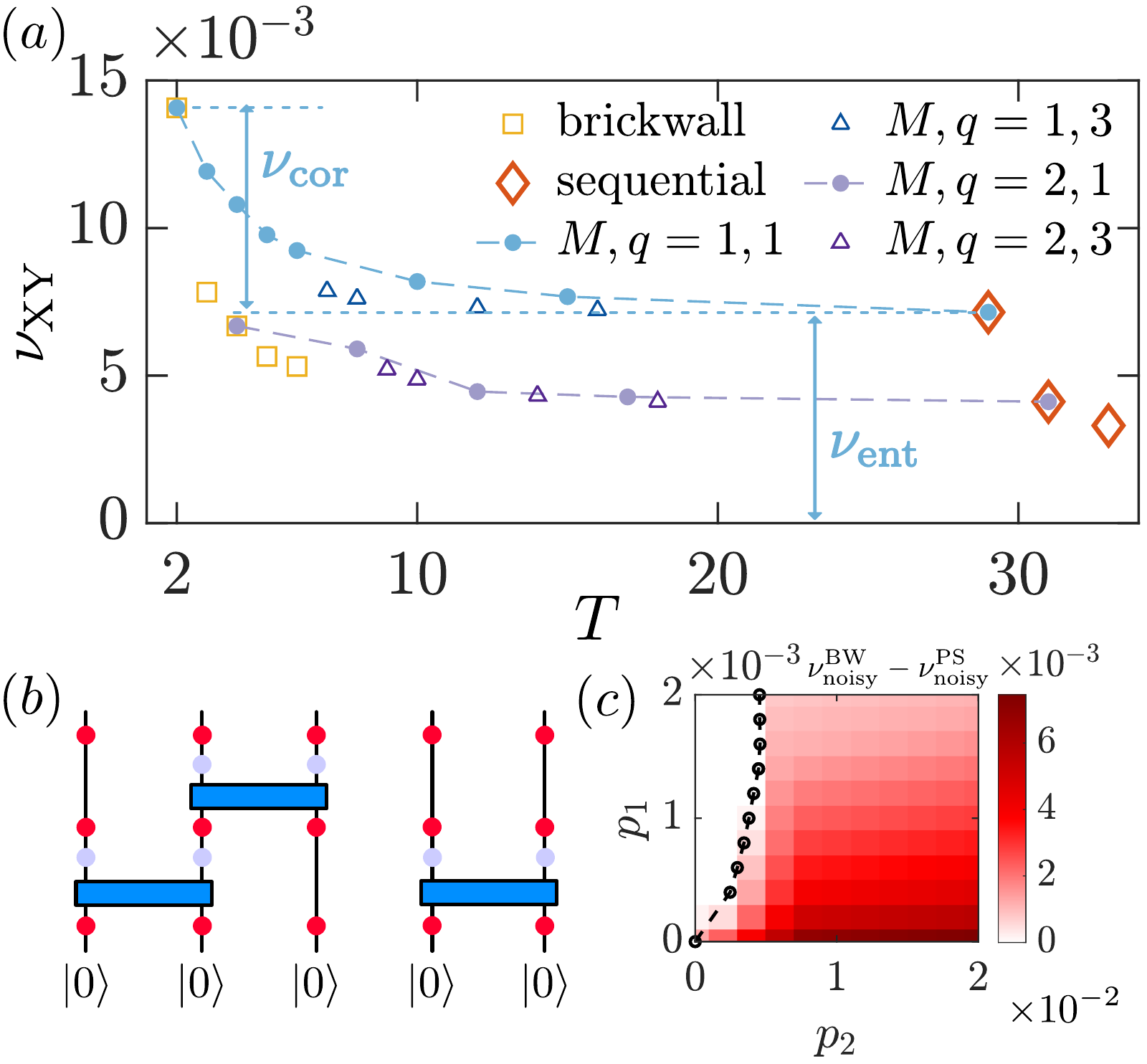}
        \caption{\textbf{Preparing the ground state of the XY model.} (a) The ground-state energy density $\nu_{\rm XY}$ of the XY model for brickwall, sequential, and various PS circuits for $M\in\{1,2\}$ (dots connected by the same line indicate tuning $l$). We visually indicate the correlation error $\nu_{\rm cor}$ and entanglement error $\nu_{\rm ent}$ for the $M=1$ case. (b) Illustration of the noise model, where red (purple) dots represent idling errors (two-qubit gate errors) applied with probabilities $p_1$ ($p_2$). (c) The difference between the optimal energy densities [\cref{err_scale}] obtained from noisy brickwall circuits, denoted as $\nu_{\rm noisy}^{\rm BW}$, and those obtained from PS circuits (which include brickwall circuits), denoted as $\nu_{\rm noisy}^{\rm PS}$, for various values of $p_1$ and $p_2$. The black dots represent the ``phase boundary'' where $\nu_{\rm noisy}^{\rm BW} = \nu_{\rm noisy}^{\rm PS}$. The system size for all plots is $N=30$.}
        \label{xy_err_fig}
\end{figure}

In \cref{xy_err_fig}(a), we plot the energy density $\nu_{\rm XY}$ along lines of constant $M\in\{1,2\}$ and for $q=1$, interpolating between brickwall and sequential circuit. As we increase the length of the sequential chunks $l$, the approximation error monotonically decreases, because the circuit is better able to capture longer ranged correlations. Increasing $M$ from $1$ (blue) to $2$ (purple) shifts the interpolation curve down uniformly, as the circuits can approximate more of the entanglement spectrum. This observation suggests decomposing the approximation error into two distinct contributions, expressed as 
\begin{equation} \label{}
\nu_{\rm XY} \approx \nu_{\rm ent} + \nu_{\rm cor},	
\end{equation}
arising respectively from the error due to lacking enough entanglement ($\nu_{\rm ent}$) and the correlation range ($\nu_{\rm cor}$), respectively. Importantly, the correlation error $\nu_{\rm cor}$ drops off rapidly in $l,q$, allowing the PS circuit to achieve an energy density $\nu_{\rm XY}$ comparable to the sequential circuit with the same number of layers $M$, while requiring a significantly smaller circuit depth [cf.~\cref{xy_err_fig}(a)]. 

\section{PS circuits on noisy devices}The low depth and sparsity of PS circuits combined with their competitive approximation error suggest that they offer an advantage in the presence of both idling and gate error. To investigate this, we consider an incoherent error model, illustrated in \cref{xy_err_fig}(b), since coherent errors can often be converted into incoherent ones using randomized compiling techniques~\cite{PhysRevX.11.041039}. In this model, single-qubit depolarizing errors act on the entire system at each time step with rate $p_1$, representing idling errors, Additionally, each two-qubit gate is followed by depolarizing errors acting simultaneously on the involved qubits at a rate $p_2$, representing two-qubit gate errors.

\subsection{Energy error scaling}

We simulate the the ground-state preparation circuits obtained in \cref{xy_err_fig}(a) in the presence of noise and compute the resulting energy $E_{\rm noisy} = {\rm Tr} (\rho_{\rm PS} H_{XY})$ and the energy density $\nu_{\rm noisy} \equiv (E_{\rm noisy} - E_{\rm GS}) / N$. For dilute noise (given by $p_1T$ and $2p_2M$), each error event perturbs only a local patch of the state. First-order perturbation theory then implies that their contributions to any extensive observable (e.g. the energy $E_{\rm noisy}$) add up linearly.
Consequently, the energy density error scales as
\begin{equation} \label{err_scale}
\nu_{\rm noisy} \approx \nu_{\rm XY} + c_E (p_1 T + 2 p_2 M),
\end{equation}
which we confirm numerically in \cref{apd_noise_E}. Here $p_1 T$ (resp. $2p_2 M$) counts the total density of idling errors (two-qubit gate errors), with a prefactor $c_E\approx 0.7$ for the XY model. For any $p_1 > 0$, the error density in sequential circuits grows linearly in system size $\nu_{\rm noisy}^{\rm SEQ} \sim c_1 p_1 N$, which make them unsuitable for large systems. Therefore, we focus on comparing brickwall and the new layouts within PS circuits in the following analysis.

Given the noise rates $p_1$ and $p_2$, we ask whether there exists a layout within the PS circuits that outperforms brickwall circuits. In the regime in which $p_2 \lesssim p_1$, this generally is not the case, as idling is just as detrimental as performing a gate.
The more interesting setting is when the gate error rate is substantially larger than the idling error rate ($p_2\gg p_1$), which is also the experimentally relevant regime~\cite{evered2023high,moses2023race,acharya2024quantum}.
In this regime, slightly increasing the circuit depth to capture longer-range correlations is inexpensive, provided the total number of gates remains unchanged. This can be precisely achieved in PS circuits by using longer sequential segments (increasing $l$). This analysis is numerically verified in \cref{xy_err_fig}(c), where we evaluate $\nu_{\rm noisy}$ for all circuits examined in \cref{xy_err_fig}(a). Specifically, we compare the minimal energy density obtained from all depths of brickwall circuits (denoted as $\nu_{\rm noisy}^{\rm BW}$) to that obtained from PS circuits (denoted as $\nu_{\rm noisy}^{\rm PS}$). Over a broad range of parameters $p_2 \gtrsim p_1 > 0$, optimized PS circuits consistently yield lower energy densities compared to brickwall circuits. This advantage diminishes as $p_1$ approaches $p_2$, delineating a kind of ``phase boundary". Note that, since brickwall circuit layouts are contained within the PS circuit layouts, in the regime $p_2 \lesssim p_1$ [the white region in \cref{xy_err_fig}(c)], the optimal PS circuit reduces to the brickwall circuit. Consequently, $\nu_{\rm noisy}^{\rm BW} - \nu_{\rm noisy}^{\rm PS} = 0$ in that region.

These results show a superior performance of PS circuits when preparing one dimensional ground states on noisy devices in a broad and relevant parameter regime. In the following, we show that their optimized layout also translates to better trainability and lower error propagation.

\subsection{Gradient variance scaling} To explore the trainability of noisy PS circuits under gradient-based optimization, we parameterize the circuit gates with parameters ${\bm \theta}$, such that the output density matrix is $\rho_{\rm PS}({\bm \theta})$ [cf.~\cref{apd_vqe_form}]. Choosing $E_{\rm noisy}$ as the cost function, we investigate the scaling of the gradient variance, 
\begin{equation} \label{}
V_E(M, T, p_1, p_2) \equiv \langle (\partial_{\theta_j} E_{\rm noisy})^2 \rangle_{\bm \theta} - \langle \partial_{\theta_j} E_{\rm noisy} \rangle^2_{\bm \theta},	
\end{equation}
where the averaging is performed over random gate realizations and all variational parameters ${\bm \theta}$. 

As $H_{XY}$ is a sum of local observables, it has been shown that $V_E^{\rm ideal} \equiv V_E^{\rm noisy}(M, T, 0, 0) \sim e^{-\alpha M}$ for noiseless circuits~\cite{cerezo2021cost,zhang2024absence}, and $V_E^{\rm noisy}(M, T, p_1, 0) \sim e^{-\alpha M - c_V p_1 T}$ for circuits under idling depolarizing noise~\cite{wang2021noise}, where $\alpha$ and $c_V$ are constant factors determined by the cost function. Building on these results, we expect that [cf.~\cref{ve_scale}]
\begin{equation} \label{ve_scale}
	V_E^{\rm noisy} (M, T, p_1, p_2) \sim e^{-\alpha M - c_V(p_1 T + 2 p_2 M)},
\end{equation}
which we numerically verify for PS circuits (including brickwall and sequential), with the cost-function dependent prefactors $\alpha \approx 0.6$ and $c_V \approx 5.5$ [cf.~\cref{apd_vqe_form}].

One consequence of this scaling is the noise-induced barren plateaus for sequential circuits~\cite{wang2021noise}, where $V_E^{\rm SEQ} \sim e^{-c_V p_1 N}$ decays exponentially with $N$. In contrast, PS circuits (including brickwall circuits) with depth $T = O(\log N)$ exhibit at most a polynomial decay of $V_E$ with system size $N$, although for a large number of layers $M$, the variance can still be small. PS circuits achieve a similar precision to brickwall circuits when approximating short-range correlated states but using fewer layers $M$ [cf.~examples in \cref{exp_pow_fig,xy_err_fig}], and thus exhibit better trainability than brickwall circuits, essentially because the PS circuits are sparser. We provide a numerical example to illustrate the behavior of $V_E$ in \cref{err_fig}(a). Together with the result shown in \cref{xy_err_fig}(c), which demonstrates that there exist PS circuits achieving a ground-state energy density lower than that of brickwall circuits with arbitrary depth, this indicates that, on noisy devices, PS circuits admit layouts that simultaneously achieve higher accuracy and better trainability than brickwall and sequential circuits.

\subsection{Error propagation}
\begin{figure}[h!]
	\centering
	\includegraphics[width=0.48\textwidth]{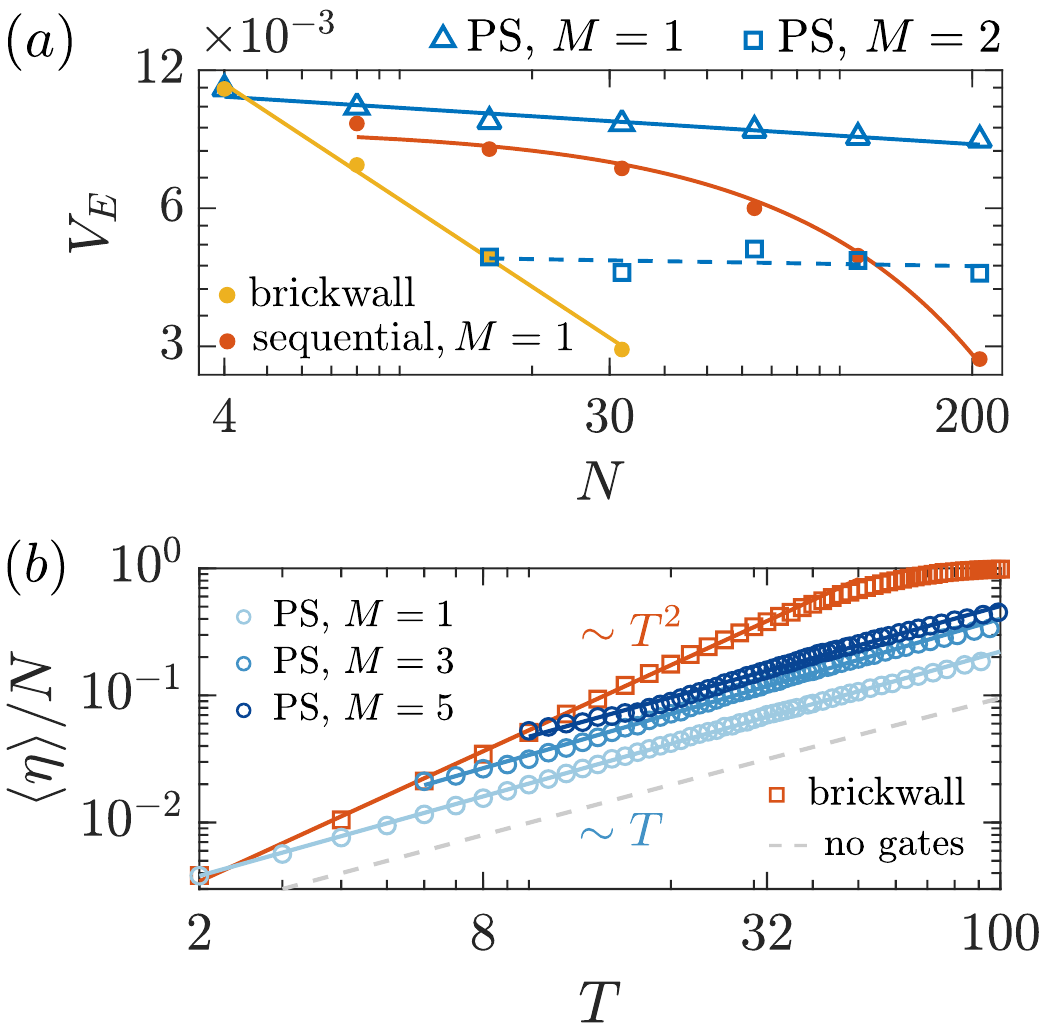}
        \caption{\textbf{Noise robustness of random PS circuits.} (a) The scaling of gradient variance $V_E$ with system size $N$ for sequential circuit ($M=1$), brickwall of depth $T = \log_2 N$, and PS circuits with the same depth, but fixed $M=1,2, q = 1$. The noise rates are $p_1 = 10^{-3}$ and $p_2 = 10^{-2}$. The lines are fits to the data.
        (b) The fraction of depolarized qubits, $\langle \eta \rangle / N$, as a function of circuit depth $T$ for brickwall circuits, PS circuits with fixed $M$ and $q=1$, and purely depolarizing dynamics with no gates. The system parameters are $N = 1000$, $p_1 = 5 \times 10^{-4}$, and $p_2 = 0$. The solid lines are fits to the data in the regime of $\langle \eta \rangle / N \ll 1$.
        } 
        \label{err_fig}
\end{figure}

Gates are known to proliferate errors in the sense that an error in a single location can lead to errors on many qubits~\cite{Nielsen2000,PRXQuantum.3.040326}.
To probe how error propagation depends on circuit layout, we consider the setting proposed in Ref.~\cite{PRXQuantum.3.040326}. Given a circuit layout, the authors consider the family of random unitaries $U$ obtained by taking each gate in the circuit to be Haar random. In the absence of errors, applying $U^\dagger U$ to $\ket0^{\otimes N}$ first entangles the qubits and then disentangles them again. If single-qubit depolarizing error occurs at a certain rate $p_1$ during this evolution, the system fails to return to its initial state and instead has some number of depolarized qubits. This process can be modeled as a Markov chain, where the system is represented as a string of zeros and ones: zeros denote noiseless qubits, and ones denote depolarized qubits [cf.~\cref{apd_ep_calc}]. Using Markov-chain Monte Carlo, the average number of depolarized qubits, $\langle \eta \rangle$ (equivalently, the average number of ones in the Markov chain), can be computed at the end of the circuit evolution~\cite{PRXQuantum.3.040326}, as shown in ~\cref{err_fig}(b).
The results in this subsection further demonstrate the noise robustness of PS circuits, albeit in a generic average case, where we can make analytical progress.

We now investigate how $\langle \eta \rangle$ depends on circuit layout in the limit of low error densities $p_1T\ll 1$. If no gates are applied, the system idles for time $2T$, and thus the depolarizing dynamics leads to $\langle \eta \rangle/N = 2p_1 T + O((p_1 T)^2)$.
Adding in gates, it can be shown that their effect when acting on a clean and a depolarized qubit amounts to depolarizing both (propagating the error) with probability 4/5 or removing the error with probability 1/5%
~\cite{PRXQuantum.3.040326}. In one-dimensional random brickwall circuits an error will thus propagate to a number of qubits that scales linearly in $T$ (for $T<N$), and thus $\langle \eta \rangle/N \propto p_1 T^2$~\cite{PRXQuantum.3.040326}, with the quadratic scaling $\propto T^2$ indicating strong error propagation.

Before we extend the analysis to general PS circuits, we first consider sequential circuits with $M$ layers (the opposite limit of brickwall circuits), with $M\ll N$. A simple argument shows that a single layer propagates an error on average over 3 additional sites, and $M$ layers produce on average a string of error of length $L(M)$, with $L(1)=4$, and $L(2)=6.4$ and asymptotically, $L(M)\to2+9M/4$ [cf.~\cref{apd_ep_calc}].
Neglecting the effect of boundaries, overlapping errors, and errors occuring on a qubit between the first and last unitary applied to it, we thus obtain (note that $T=N$ here) $\langle \eta \rangle/N = p_1 T(1+L(M))$ [cf.~\cref{apd_ep_calc}].

In general PS circuits, we numerically find that $\langle \eta \rangle$ interpolates between the brickwall and sequential limits and find that it is well approximated as
\begin{equation}\label{ep_eq}
    \langle\eta\rangle/N \approx p_1 T (c_{\eta,1} + c_{\eta,2} M),
\end{equation}
with prefactors $c_{\eta,1},c_{\eta,2}$ of order unity and exhibit a mild dependence on the circuit layout (as well as $T$ and $M$),  but only very weakly on system size [cf.~\cref{apd_ep_calc}].
Importantly, we always find the dominant contribution (in the limit of sparse errors) to grow with $p_1TM$ (see \cref{err_fig}(b) and \cref{apd_ep_calc}).
Thus, in PS circuits with constant $M$, the errors are just \emph{linear} in $T$, as compared to the $T^2$ scaling in brickwall circuits.

\section{Parallel-sequential circuits in higher dimensions}
\label{apd_hdim}
In higher-dimensional lattices, various types of sequential and brickwall circuits have been extensively studied. These include applications such as the preparation of long-range entangled states using sequential circuits~\cite{Banuls2008,zypp,yj_circ,chen2024sequential} and the demonstration of quantum computational advantage with deep brickwall circuits~\cite{Boixo2018,Arute2019a,brandao2021models,hangleiter2023computational}.
Extending the one-dimensional lattice analysis, one can define PS circuits in arbitrary higher dimensions that interpolate between higher-dimensional sequential circuits and brickwall circuits. Due to the freedom of defining the sequential circuits and the brickwall circuits, such an interpolation is not unique. For concreteness, here we consider the multi-layer sequentially-generated states (SGS)~\cite{Banuls2008}~\footnote{Here the multi-layer SGS means that we replace the arbitrary-bond-dimension sequential circuits used in Ref.~\cite{Banuls2008} by multiple layers of sequential circuits consisting of two-qubit gates.} as the representative sequential circuit and a particular ordering of applying the commuting brickwall layers, which will be detailed later. In the following, we define PS circuits on the two-dimensional square lattice, while the generalization to arbitrary-dimensional cubic lattices is straightforward.

\begin{figure}[h!]
	\centering
	\includegraphics[width=0.48\textwidth]{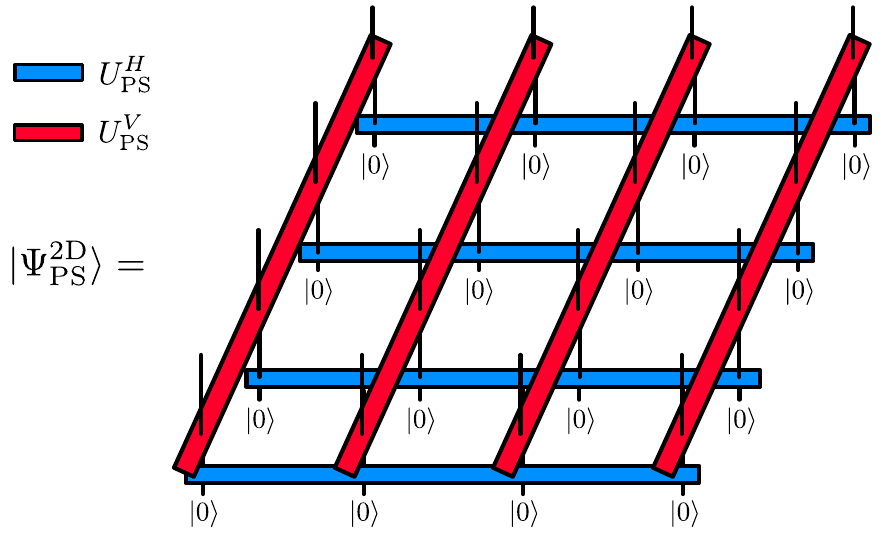}
        \caption{Two-dimensional parallel-sequential circuits, $U_{\rm PS}^{\rm 2D}$, and their associated variational states, $| {\Psi^{2D}_{{\rm PS}}} \rangle$, are constructed by consecutively applying `lines' of 1D PS circuits $U_{\rm PS}^H, U_{\rm PS}^V$ along both lattice directions to entangle the entire lattice [cf.~\cref{ps2d}], inspired by the construction of the sequentially-generated states~\cite{Banuls2008}.}
        \label{psc_2d}
\end{figure}

Consider a two-dimensional square lattice of size $N=L \times L$ [c.f.~\cref{psc_2d}]. We define the PS \textit{unitary} acting on the $j$-th row of the 2D lattice as $U_{\rm PS}^{H}(j,{\bm \theta},L,M,l,q)$, where the gate layout within the unitary follows that of the 1D PS circuit. Here ${\bm \theta}$ denote the gate parameters, and the circuit layout is characterized by the system linear size $L$, number of gate layers $M$, the chunk length $l\geq 2$, and the overlapping distance $1\leq q\leq l$ [cf.~\cref{psc_illu,ps_state}]. Similarly, we define the PS unitary acting on the $k$-th column of the 2D lattice as $U_{\rm PS}^{V}(k,{\bm \theta},L,M,l,q)$.

Inspired by the construction of SGS~\cite{Banuls2008}, we construct the 2D PS circuits by first applying row unitaries, followed by column unitaries, to cover the 2D lattice (as illustrated in \cref{psc_2d}), as
\begin{widetext}
\begin{equation} \label{ps2d}
U_{\rm PS}^{\rm 2D}({\bm \theta}_{\rm 2D},L,M,l,q) = 	\prod\limits_{k = 1}^{L} U_{\rm PS}^{V}(k,{\bm \theta}_{V,k},L,M,l,q)  \prod\limits_{j = 1}^{L} U_{\rm PS}^{H}(j,{\bm \theta}_{H,j},L,M,l,q).
\end{equation}
\end{widetext}
Here, ${\bm \theta}_{\rm 2D} \equiv (\{{\bm \theta}_{V,k}\}, \{{\bm \theta}_{H,j}\})$ represents all gate parameters in the 2D PS circuits. The circuit depth is given by $T = 2(l + q + 2M - 3)$, and the variational states associated with the 2D PS circuits are expressed as 
\begin{equation} \label{}
| {\Psi^{2D}_{{\rm PS}}} \rangle = U_{\rm PS}^{\rm 2D}({\bm \theta}_{\rm 2D}, L, M, l, q) |0\rangle^{\otimes N}.
\end{equation}

Similar to the 1D case, the number of layers $M$ controls the amount of entanglement in the 2D PS circuits, such that the entanglement entropy $S$ of $| {\Psi^{2D}_{{\rm PS}}} \rangle$ across any horizontal or vertical cut (not in the overlapping regions) satisfies $S \leq 2ML$. By fixing $M$ and $q = 1$ while varying $l$, the 2D PS circuits interpolate between the multi-layer SGS ($l = L - 1$) and the 2D brickwall circuits ($l = 2$), which consist of multiple commuting layers of brickwall gates acting first along the horizontal direction, followed by the vertical direction. This analysis generalizes directly to arbitrary higher-dimensional cubic lattices, where the PS circuits are constructed by consecutively applying `lines' of 1D PS circuits (as defined in the main text) along all lattice directions to entangle the entire lattice.

PS circuits in higher dimensions exhibit several interesting properties. For instance, $| {\Psi^{2D}_{{\rm PS}}} \rangle$, created by single-layer ($M = 1$) PS circuits, already includes cluster states in arbitrary dimensions, which serve as universal resources for measurement-based quantum computations in two- and higher-dimensional lattices~\cite{briegel2009measurement}. By choosing $M = O(1)$ and $l, q = O(\log N)$, the PS circuits are capable of generating states with area-law entanglement and exponentially decaying correlations, making it a potential circuit model for gapped ground states in higher dimensions~\cite{hastings2006spectral}. Furthermore, constant-layer 2D PS circuits enable the efficient computation of few-body correlators using the methods outlined in Ref.~\cite{Banuls2008,zypp}. Finally, following the arguments presented in the main text, we anticipate that higher-dimensional noisy PS circuits will also suppress error proliferation and demonstrate superior trainability and evaluation accuracy compared to their higher-dimensional noisy sequential circuit and noisy brickwall circuit counterparts.

\section{Discussion and Outlook} Parallel-sequential (PS) circuits interpolate between brickwall and sequential circuits and thus constitute a new class of circuit layouts that are simultaneously expressive and robust to noise. These layouts enable the discovery of better circuits for ground state preparation. Our work suggests a versatile, easy-to-implement strategy for enhancing the performance of quantum tasks on noisy devices by replacing sequential or brickwall circuits by the optimal layout within the PS family.

Although our discussion has focused on one-dimensional systems, we expect that many of our results extend to higher-dimensional PS circuits, which can be defined analogously [cf.~\cref{apd_hdim}]. Future work should explore the properties of PS circuits, such as their information dynamics~\cite{bensa2021fastest}, simulation complexity~\cite{hangleiter2023computational}, and their applications in variational quantum algorithms~\cite{cerezo2021variational}. Additionally, PS circuits can serve as a `skeleton' for applying adaptive circuit growth techniques~\cite{grimsley2019} to identify more optimal circuit layouts for representing quantum states. It is also worth investigating the use of PS circuits to enhance numerous protocols~\cite{Ran2020a,rudolph2023decomposition,iqbal2022preentangling,gundlapalli2022deterministic,melnikov2023quantum,iaconis2024quantum,bohun2024scalable,gonzalez2024efficient,sano2024quantum} that rely on representing MPS of bond dimension $D=2$ as a single-layer sequential circuit.

\section{Acknowledgments}We thank Ignacio Cirac, Guillermo Gonz\'alez-Garc\'ia, Khadijeh Sona Najafi, Adam Smith, and Georgios Styliaris for insightful discussions. The research is part of the Munich Quantum Valley, which is supported by the Bavarian state government with funds from the Hightech Agenda Bayern Plus. ZYW acknowledge funding from the German Federal Ministry of Education and Research (BMBF) through EQUAHUMO (Grant No. 13N16066) within the funding program Quantum Technologies—From Basic Research to Market. ZYW also acknowledges support from the U.S.~Department of Energy,
Office of Science, Accelerated Research in Quantum Computing, Fundamental Algorithmic Research toward Quantum Utility (FAR-Qu). DM acknowledges support from Novo Nordisk Fonden under grant numbers NNF22OC0071934 and NNF20OC0059939.
The numerical calculations were performed using the ITensor.jl~\cite{itensor} and PastaQ.jl~\cite{pastaq}.

\appendix

\renewcommand{\thefigure}{A\arabic{figure}}
\renewcommand{\theequation}{A\arabic{equation}}
\setcounter{equation}{0}
\setcounter{figure}{0}

\section{Brief review of Matrix product state (MPS)}
\label{apd_mps_review}
\begin{figure*}
	\centering
	\includegraphics[width=0.98\textwidth]{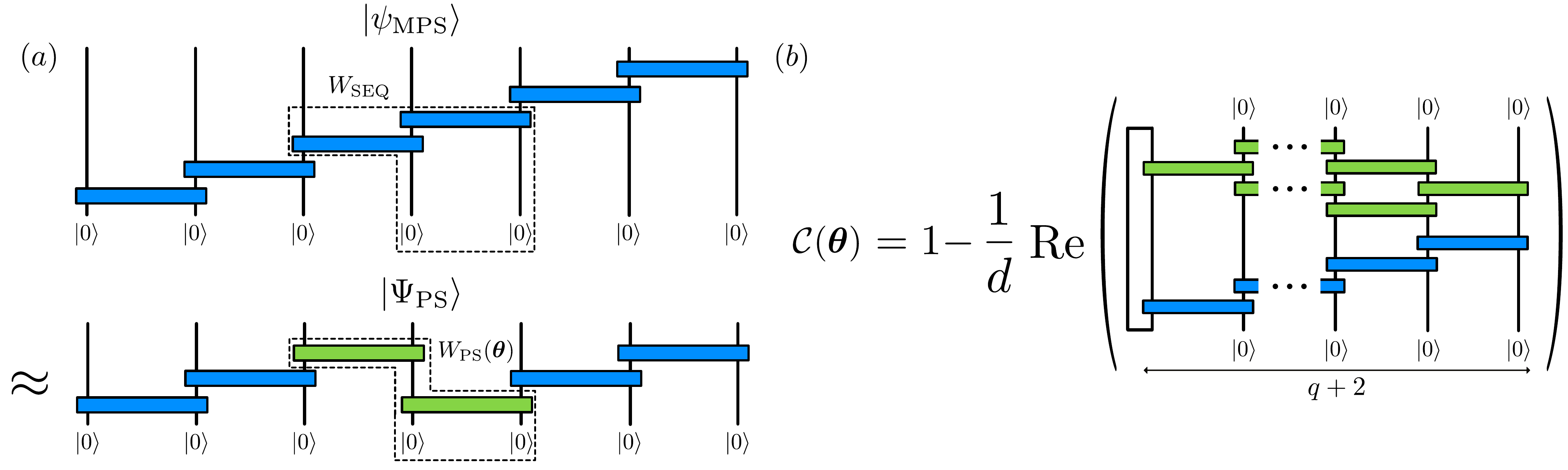}
        \caption{(a) The target MPS $|\phi_{\mathrm{MPS}}\rangle$ with bond dimension $D = d = 2$ [\cref{mps_form}] can be exactly represented as a single-layer sequential circuit composed of two-qubit gates. By finding a local isometry $W_{\mathrm{PS}}(\bm{\theta})$ (denoted by the dashed box on the right side) that approximates the isometry $W_{\mathrm{SEQ}}$ (denoted by the dashed box on the left side) in the sequential circuit, one can invert the order of gates within $W_{\mathrm{SEQ}}$, obtaining an approximation $|\Psi_{\mathrm{PS}}(\bm{\theta})\rangle \approx |\phi_{\mathrm{MPS}}\rangle$ using a PS circuit with one additional sequential chunk and reduced circuit depth. (b) Graphical representation of the cost function $\mathcal{C}(\bm{\theta})$ [\cref{optm_cost}] for general PS circuit with overlapping distance $q$.}
        \label{ps_optm_fig}
\end{figure*}

Consider a one-dimensional chain of $N$ qudits, each of dimension $d$ (we focus on qubit case $d=2$ in this work), with open boundary conditions. An MPS on this chain is expressed as:
\begin{equation}\label{mps_form}
|\phi_{\rm MPS}\rangle = \sum_{i_{1}, \ldots ,i_{N}=0} ^{d-1} A_{1}^{i_{1}} \ldots A_{N}^{i_{N}}|i_{1} \ldots i_{N}\rangle,
\end{equation}
where $\{A^{i_j}_{[j]}\}$ in the bulk are matrices of dimension $D \times D$, with $D$ denoting the bond dimension of the MPS. The boundary tensors $A_{1}^{i_{1}}$ and $(A_{N}^{i_{N}})$ are of dimensions $1\times D$ and $D\times 1$, respectively. Every MPS can be transformed into a canonical form~\cite{schollwock2011density}, where the tensors satisfy the condition $\sum_{i_k,\alpha}  (A^{*i_k}_{k})_{\alpha,\beta'} (A^{i_k}_{k})_{\alpha,\beta} =\delta_{\beta,\beta'}$. An MPS in canonical form can be represented exactly by a single \textit{dense} ``layer'' of a sequential circuit~\cite{Schon2005,Schon2007}. For example, MPS with $D = d = 2$ can be prepared by such a ``single-layer'' sequential circuit composed of two-qubit gates, as illustrated in \cref{ps_optm_fig}(a). Multi-layer sequential circuits are circuits in which a single-layer circuit composed of two-qubit gates is applied multiple times~\cite{Ran2020a,lin2021real}.

In \cref{exp_pow_fig}, we focus on cases where the MPS tensors in the bulk are identical, resulting in states that are approximately translational invariant for sufficiently large system sizes. Such MPS are referred to as \textit{bulk-TI} MPS. Specifically, we denote the tensors in the bulk as $A$, using a graphical notation
${(A^i)_{jk} =
\begin{array}{c}
    \begin{tikzpicture}[scale=.4, baseline={([yshift=-5.5ex]current bounding box.center)}, thick]
    	\ATensor{0,0}{$A$}
    	\draw (0,1.35) node {\scriptsize $i$};
    	\draw (1.35,0) node {\scriptsize $k$};
    	\draw (-1.35,0) node {\scriptsize $j$};
    \end{tikzpicture}
\end{array}},$ and the state is denoted as
\begin{equation}
|\phi_{\rm MPS}^{\textrm{bulk-TI}}\rangle
	= 
  	\begin{array}{c}
	\begin{tikzpicture}[scale=.45, baseline={([yshift=0ex]current bounding box.center)}, thick]
\draw (-1,0) -- (-1,1);
		\draw[shift={(0,0)},dotted] (0,0) -- (5,0);
		\ATensor{0,0}{$A$}
		\ATensor{1.5,0}{$A$}
		\ATensor{4.5,0}{$A$}
		\ATensor{6,0}{$A$}
\draw (7,0) -- (7,1);
	\end{tikzpicture}
	\end{array}
\end{equation}
where the boundary tensors are taken to be the identity. We also define the \textit{transfer matrix} associated with $A$ as
\begin{equation}
	E_{AA} = \sum_{i=1}^d (A^{i})^* \otimes A^i
	=
  	\begin{array}{c}
		\begin{tikzpicture}[scale=.5,thick,baseline={([yshift=1ex]current bounding box.center)}]
			\ATensor{0,0}{$A$}
			\ADaggTensor{0,1.7}{$A^*$}
		\end{tikzpicture}
	\end{array}.
	\label{eq:transfer_matrix}
\end{equation}

The correlations within MPS are characterized by the two-body correlation function:
\begin{equation} \label{cor_func}
C(i,j) = \langle O_i O_j \rangle - \langle O_i \rangle \langle O_j \rangle,	
\end{equation}
where $O_i$ and $O_j$ are single-site operators acting on sites $i$ and $j$, respectively. For bulk-TI MPS of $N\gg 1$, the correlation function follows a translationally invariant form in the bulk, $C(i,j) = C(|i-j|)$, which can be derived from the transfer matrix of the MPS~\cite{Wolf2006}:
\begin{equation} \label{cor_func_decay}
C(r) = \sum_{j=1}^{D^2-1} C_j \exp(-r/\xi_j),
\end{equation}
where $C_1, C_2, \ldots, C_{D^2-1}$ are constants, and $\xi \geq \xi_2 \geq \cdots \geq \xi_{D^2-1} > 0$ represent the length scale of each decay. The largest $\xi$ is referred to as the correlation length of the state and characterizes the asymptotic exponential decay of correlations.

In \cref{exp_pow_fig}, we examine a bulk-TI MPS family~\cite{Wolf2006} and random bulk-TI MPS with bond dimension $D = 2$. The bulk tensors for the bulk-TI MPS family are given by [cf.~\cref{mps_form}]:
\begin{equation} \label{cls_mps}
A_{[i]}^{0}(g) = \left(\begin{array}{cc}
0 & 0 \\
1 & 1
\end{array}\right), \quad A_{[i]}^{1}(g) = \left(\begin{array}{cc}
1 & g \\ 
0 & 0
\end{array}\right),
\end{equation}
where $g \in (-1,0)$ interpolates the state from the cluster state ($g = -1$) to the GHZ state ($g = 0$). In this regime, the eigenvalues of the transfer matrix are $\{\lambda_i\} = \{1, \frac{1+g}{1-g}, 0, 0\}$. Consequently, the correlation function $C(r)$ decays as a single exponential [cf.~\cref{cor_func}], with the correlation length given by
\begin{equation} \label{qpt_cor_g}
\xi = \left(\ln \frac{1 - g}{1 + g} \right)^{-1}.
\end{equation}
By tuning $g$, we can construct MPS of varying correlation lengths. Note that $g \in (-1,0)$ encompasses all states with $g < 0$, as the tensors $\{A(g)\}$ in \cref{cls_mps} can be mapped to $\{A(1/g)\}$ through a gauge transformation~\cite{Wolf2006,zyadi}. For the random bulk-TI MPS with $D = 2$, we sample the bulk tensor from the Haar random distribution~\cite{garnerone2010statistical,haferkamp2021emergent}.

\section{PS circuits for quantum state preparation}
In this section, we discuss various aspects of PS circuits for quantum state preparation, including the optimization method used to approximate MPS with $D = 2$, their performance analysis, and a comparison with the log-depth RG circuit introduced in Ref.~\cite{malz2023preparation}.

\subsection{Approximating MPS of $D=2$}

\subsubsection{Local optimization algorithm}

\label{algo_mps}

Given a single-layer sequential circuit corresponding to the target $|\phi_{\mathrm{MPS}}\rangle$ [cf.~\cref{mps_form}] with bond dimension $D=2$, we propose an efficient local optimization method to obtain the PS state $|\Psi_{\mathrm{PS}}\rangle$ that best approximates $|\phi_{\mathrm{MPS}}\rangle$. This approach is illustrated in \cref{ps_optm_fig}(a), where it is shown that a sequential circuit can be parallelized by locally inverting the order of gates in a local isometry $W_{\mathrm{SEQ}}
	= 
  	\begin{array}{c}
	\begin{tikzpicture}[scale=.45, baseline={([yshift=0ex]current bounding box.center)}, thick]
\draw[shift={(0,0)},dotted] (-1.5,0) -- (-1,0);
		\draw[shift={(0,0)},dotted] (0,0) -- (3,0);
		\ATensor{0,0}{$A$}
		\ATensor{3,0}{$A$}
		\draw [decorate,
    	decoration = {calligraphic brace,mirror}] (0,-0.8) --  (3,-0.8);
		\draw (1.5,-1.5) node {\scriptsize $q + 1$};
	\end{tikzpicture}
	\end{array}$
, resulting in another isometry $W_{\mathrm{PS}}(\bm{\theta})$. This local gate-order inversion leads to a PS circuit with one additional sequential chunk. While an exact solution ${{\bm \theta}_0}$ satisfying $W_{\mathrm{PS}}({{\bm \theta}_0}) = W_{\mathrm{SEQ}}$ typically does not exist, we can numerically minimize the distance between $W_{\mathrm{PS}}(\bm{\theta})$ and $W_{\mathrm{SEQ}}(\bm{\theta})$ using the following cost function
\begin{align}
\mathcal{C}(\bm{\theta}) 
&= \frac{1}{2d} \|W_{\rm PS} - W_{\rm SEQ} \|_F^2 \nonumber \\
&= 1 - \frac{1}{d} \operatorname{Re}[\operatorname{Tr}(W_{\mathrm{PS}}^{\dagger}(\bm{\theta}) W_{\mathrm{SEQ}})]
\label{optm_cost}
\end{align}
to find the optimal parameters ${\bm{\theta}}_{\rm{best}}$, where $\|A\|_F = \sqrt{\operatorname{Tr}(A^\dagger A)}$ is the Frobenius norm. Figure \ref{ps_optm_fig}(b) graphically depicts $\mathcal{C}(\bm{\theta})$ for generic PS circuits with overlapping distance $q$. Since this local optimization involves only $O(q)$ gates, it can be efficiently solved using a DMRG-like sweep~\cite{evenbly2009algorithms}. This optimization algorithm can be employed to prepare general MPS with bond dimension $D=2$. To achieve this, it needs to be applied to multiple regions of the 1D chain to construct the PS circuit that approximately prepares the target MPS. In the following, we focus on bulk-TI MPS, where the same solution $W_{\mathrm{PS}}({\bm{\theta}}_{\mathrm{best}})$ can be used to replace $W_{\mathrm{SEQ}}(\bm{\theta})$ in different regions of the chain. For example, one can apply this replacement to create a PS circuit of chunk length $l$ (with $n_C = \lceil N/l\rceil$ chunks) and overlapping distance $q$, with its structure graphically represented as
\begin{widetext}
\begin{equation}
|\Psi_{\rm PS}\rangle
	= 
  	\begin{array}{c}
	\begin{tikzpicture}[scale=.45, baseline={([yshift=0ex]current bounding box.center)}, thick]
\draw[shift={(0,0)},dotted] (-1.5,0) -- (-1,0);
		\draw[shift={(0,0)},dotted] (0,0) -- (3,0);
		\ATensor{0,0}{$A$}
		\ATensor{3,0}{$A$}
		\draw [decorate,
    	decoration = {calligraphic brace,mirror}] (0,-0.8) --  (3,-0.8);
		\draw (1.5,-1.5) node {\scriptsize $l - q - 1$};
		\WTensor{5,0}{$W_{\rm PS}$}
		\draw [decorate,
    	decoration = {calligraphic brace,mirror}] (4,-0.8) --  (6,-0.8);
		\draw (5,-1.5) node {\scriptsize $q+1$};
		\draw[shift={(0,0)},dotted] (7,0) -- (10,0);
		\ATensor{7,0}{$A$}
		\ATensor{10,0}{$A$}
				\draw [decorate,
    	decoration = {calligraphic brace,mirror}] (7,-0.8) --  (10,-0.8);
		\draw (8.5,-1.5) node {\scriptsize $l - q - 1$};
		\WTensor{12,0}{$W_{\rm PS}$}
		\draw [decorate,
    	decoration = {calligraphic brace,mirror}] (11,-0.8) --  (13,-0.8);
		\draw (12,-1.5) node {\scriptsize $q+1$};
		\draw[shift={(0,0)},dotted] (14,0) -- (17,0);
		\ATensor{14,0}{$A$}
		\ATensor{17,0}{$A$}
						\draw [decorate,
    	decoration = {calligraphic brace,mirror}] (14,-0.8) --  (17,-0.8);
		\draw (15.5,-1.5) node {\scriptsize $l - q - 1$};
		\draw[shift={(0,0)},dotted] (18,0) -- (18.5,0);
	\end{tikzpicture}
	\end{array}
\end{equation}
\end{widetext}

\begin{figure*}
	\centering
	\includegraphics[width=0.68\textwidth]{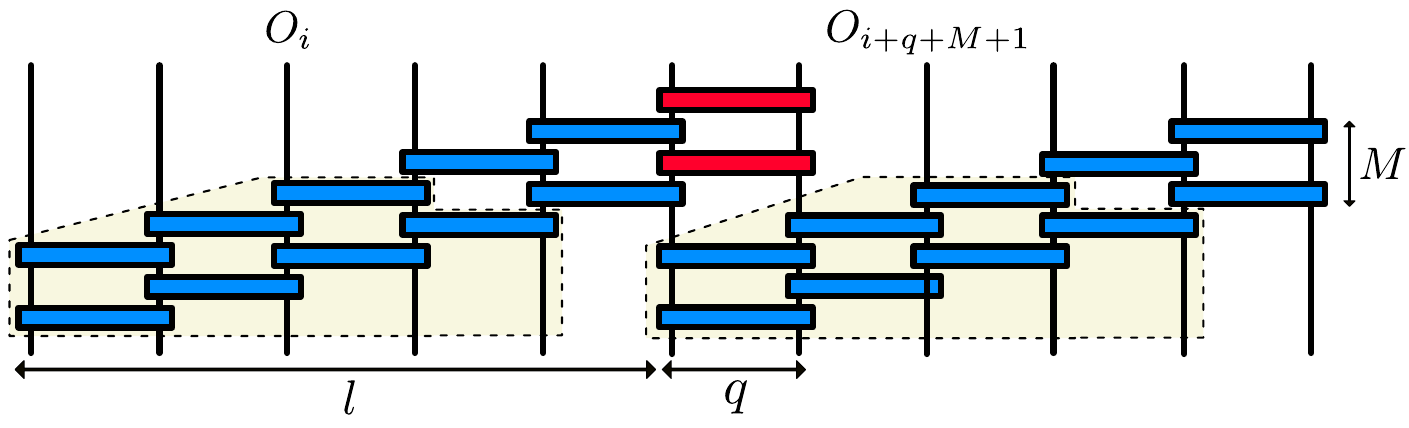}
        \caption{The correlation structure in PS circuits. Sites within each chunk (of length $l$) are correlated with each other, whereas correlations between sites in neighboring chunks are more limited. As an example, the regions within the dashed boxes illustrate the inverse light cones of site $i$ and site $i+q+M+1$ near the overlap region, with $q=2$ and $M=2$ shown here. Since the inverse light cones do not overlap, no correlation exists between these two sites.}
        \label{ps_cor_illu}
\end{figure*}

\subsubsection{Error due to missing correlations, and the scaling of the fidelity $\cal F$ for bulk-TI MPS}
\label{apd_cor_fid}

The gate order inversion [cf.~\cref{ps_optm_fig}] cannot be realized perfectly due to the inherent light-cone structure of the PS circuits. This is illustrated in \cref{ps_cor_illu}, where sites $i$ and $i+q+M+1$ ($M=1$ here for MPS of $D=2$) are not correlated, i.e., $C(i,i+q+M+1) = 0$ [cf.~\cref{cor_func}] for arbitrary local operators. As a result, approximation errors arise when attempting to use PS circuits to capture states with correlations beyond the distance ${\cal R}_c = q+M$, such as gapped ground states with exponentially decaying correlations. These errors can be reduced by increasing $q$, as shown in \cref{exp_pow_fig}(a).

We can quantify this error due to the missed correlation by studying the state fidelity between the target bulk-TI MPS $|\phi_{\rm MPS}^{\textrm{bulk-TI}}\rangle$ with bond dimension $D = 2$ and its best approximation $|\Psi_{\rm PS}({\bm \theta}_{\rm best})\rangle$ within a given PS circuit, defined as ${\cal F} = |\langle \Psi_{\rm PS}({\bm \theta}_{\rm best}) | \phi_{\rm MPS}^{\textrm{bulk-TI}}\rangle|^2$. To compute ${\cal F}$, we define the overlap matrix between $W_{\rm PS}$ and $W_{\rm SEQ}$ as $
    E_{AW} = 
		\begin{tikzpicture}[scale=.39,thick,baseline={([yshift=-0.5ex]current bounding box.center)}]
			\WTensor{0,0}{$W_{\rm SEQ}$}
			\PDaggTensor{0,1.5}{W_{\rm PS}^*}
		\end{tikzpicture}$
, and the fidelity $\cal F$ can be expressed as
\begin{align} \label{f_form}
{\cal F} &=	|\langle \Psi_{\rm PS} |\phi_{\rm MPS}^{\textrm{bulk-TI}}\rangle|^2  \nonumber \\&= |\langle I| (E_{AA}^{l-q-1}E_{AW})^{n_C-1} E_{AA}^{l-q-1} |I\rangle|^2,
\end{align}
where $\langle I|$ and $|I\rangle$ denote the left and right boundary conditions. This already implies the exponential decay of $\cal F$ with the number of sequential chunks $n_C$ (i.e. the times that we replace $W_{\rm SEQ}$ with $W_{\rm PS}$), as
\begin{equation} \label{}
{\cal F} \approx \exp[-\kappa(q) \cdot (n_C-1)],	
\end{equation}
where $\kappa(q)$ represents the error rate. This scaling behavior is numerically verified for all bulk-TI MPS analyzed in this work (see \cref{scale_supp}(a) for numerical results corresponding to the MPS family discussed in \cref{cls_mps}). The extracted numerical scaling is utilized to determine $\kappa(q)$, as shown in \cref{exp_pow_fig}(a).

Moreover, for $l - q - 1 \gg 1$, the power of $E_{AA}$ can be approximated as $E_{AA}^{l-q-1} \approx |\Lambda_L\rangle \langle \Lambda_R|$, where $|\Lambda_L\rangle$ and $|\Lambda_R\rangle$ denote the left and right eigenvectors associated with the eigenvalue 1. The normalization condition implies 
\begin{equation} \label{}
\langle I | \Lambda_L\rangle \langle \Lambda_R | I\rangle \approx \langle\phi_{\rm MPS}^{\textrm{bulk-TI}}|\phi_{\rm MPS}^{\textrm{bulk-TI}}\rangle = 1.	
\end{equation}
Consequently, in this regime, the fidelity can be expressed as
\begin{align}
\mathcal{F} 
&\approx |\langle I|\Lambda_L\rangle \langle \Lambda_R|I\rangle 
         (\langle \Lambda_R| E_{AW} |\Lambda_L\rangle)^{n_C - 1}|^2 \nonumber \\
&\approx |\langle \Lambda_R| E_{AW} |\Lambda_L\rangle|^{2(n_C - 1)} 
\label{fscale}
\end{align}
which provide another way to extract $\kappa(q)\approx  -\log (|\langle \Lambda_R| E_{AW} |\Lambda_L\rangle|^2)$.

\subsubsection{Details on CNOT depth scaling for $D=2$ bulk-TI MPS [~\cref{exp_pow_fig}(b)]}
\label{cnot_scale_sec}
The scaling of fidelity $\cal F$ [\cref{scale_supp}(a) and \cref{fscale}] and the error density $\kappa(q) = \kappa_0 \exp(-\gamma q / \xi)$ [cf.~\cref{exp_pow_fig}(a)] allow us to deduce the optimal PS circuit and the scaling of its parameters for preparing the target $D=2$ MPS of size $N$ with a given infidelity $\epsilon = 1 - \cal F$.

\begin{figure*}
	\centering
	\includegraphics[width=0.8\textwidth]{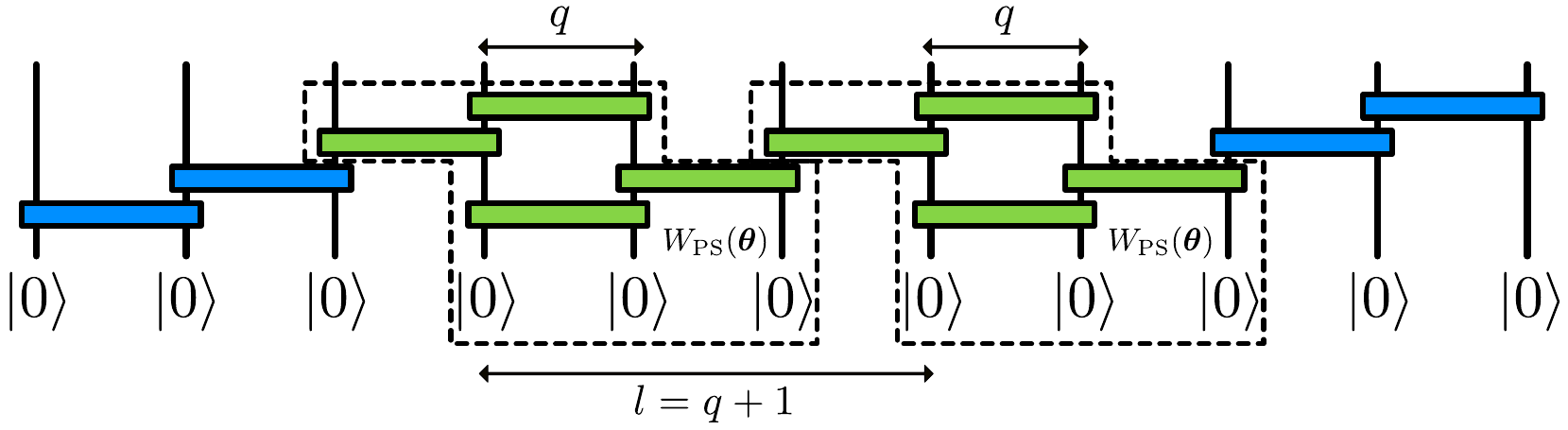}
        \caption{Illustration of the single-layer PS circuit with  $l = q + 1$, representing the maximal level of parallelization (i.e., the minimal value of  $l$ ) at which the local optimization algorithm remains applicable.}
        \label{psc_max_para}
\end{figure*}

From \cref{fscale}, for a given $q$ and target infidelity $\epsilon \ll 1$, the maximum number of sequential chunks should be approximately \( n_C \approx \epsilon / \kappa(q) + 1 \). Since \( n_C \approx N / l \) and the circuit depth for single-layer PS circuits is \( T = l + q - 1 \), we choose \( l = q + 1 \) to minimize \( T \). This choice corresponds to the minimal \( l \) that allows our optimization algorithm (which locally replaces \( W_{\rm SEQ} \) with \( W_{\rm PS} \)) to function effectively, as illustrated in [cf.~\cref{psc_max_para}]. Combining these considerations, we obtain the relation
\begin{equation}
 \frac{N}{q+1} \approx \frac{\epsilon}{\kappa_0} \exp(\gamma q / \xi) + 1,
\end{equation}
from which we deduce the asymptotic scaling of \( q \) (and thus \( l \)) as \( q \sim \xi \cdot \log (\kappa_0 N / \epsilon) \), and the scaling of circuit depth as \( T = 2q \sim \xi \cdot \log (\kappa_0 N / \epsilon) \).

To estimate the CNOT depth, we use the theoretical lower bound for implementing a generic \( m \)-to-\( n \) qubit isometry~\cite{PhysRevA.93.032318}:
\begin{equation} \label{iso_tcnot}
	T_{\text {iso}}(m, n) = \left\lceil \frac{1}{4} \left(2^{n+m+1} - 2^{2m} - 2n - m - 1\right) \right\rceil.
\end{equation}
In some cases, this bound is tight~\cite{rakyta2022approaching}. For the first layer of gates acting on the initial qubits, we have \( T_{\text {iso}}(0, 2) = 1 \), and for generic two-qubit gates, \( T_{\text {iso}}(2, 2) = 3 \). Thus, the CNOT depth is \( T_{\rm CNOT} = 3T \sim \xi \cdot \log (\kappa_0 N / \epsilon) \), as numerically demonstrated in \cref{exp_pow_fig}(b).

For comparison, we also show \( T_{\rm CNOT} \) for the sequential-RG circuit described in Ref.~\cite{malz2023preparation} for preparing the same set of \( D=2 \) MPS, in \cref{exp_pow_fig}(b). The details of the \( T_{\rm CNOT} \) calculation are provided in the supplementary materials of Ref.~\cite{malz2023preparation}. We find that PS circuits consistently yield shallower circuits for preparing the same state for a given system size \( N \). This difference arises from the multi-qubit isometries used in RG circuits~\cite{malz2023preparation}, such as the 2-to-3 isometries (with \( T_{\text {iso}}(2, 3) = 14 \)) employed in sequential-RG circuits for \( D=2 \) MPS. In \cref{ps_rg_compare}, we show that the advantage of PS circuits over the RG circuits persists for states with higher bond dimensions.

\subsubsection{Robustness against local minima}
\label{apd_lm}

To determine the gate parameters ${\bm \theta}_{\rm best}$ of the PS circuit for preparing the bulk-TI MPS $|\phi _{\rm MPS}^{\rm bulk-TI}\rangle$ of size \( N \), correlation length \( \xi \), and infidelity \( \epsilon \), it suffices to find the global minimum of the cost function \( {\cal C}(\bm \theta) \) [cf.~\cref{ps_optm_fig}(c)] for a one-dimensional local structure characterized by the overlapping distance \( q \sim \xi \cdot \log (N / \epsilon) \) [cf.~\cref{cnot_scale_sec}]. This ensures computational efficiency for each run of the DMRG-like sweep~\cite{evenbly2009algorithms} inside $W_{\rm PS}(\bm \theta)$. However, optimizations on quantum circuits are known to suffer from the presence of local minima in the optimization landscape~\cite{cerezo2021variational,bharti2022noisy}, raising concerns about whether our algorithm can asymptotically and efficiently find \( {\bm \theta}_{\rm best} \) corresponding to the globally minimal fidelity error density \( \kappa(q) \).

\begin{figure}[h!]
	\centering
	\includegraphics[width=0.48\textwidth]{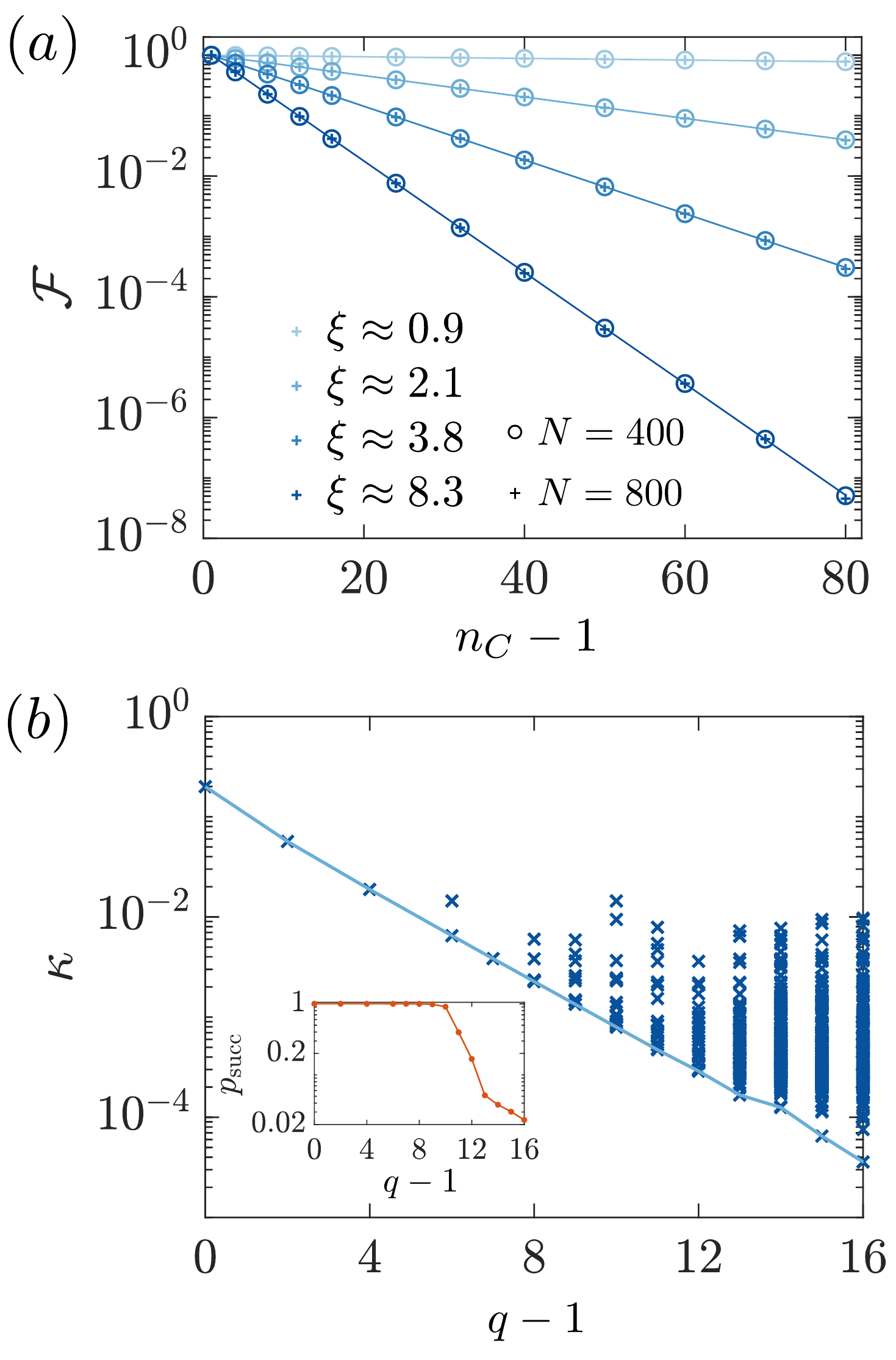}
        \caption{(a) The scaling of fidelity $\cal F$ [\cref{f_form}] with the number of chunks $n_C$ for various states in the MPS family [cf.~\cref{cls_mps}] and system sizes. Here we fix the overlapping distance $q=2$, and the chunk length $l$ is varied to get different $n_C$. (b) The error density $\kappa(q)$ achieved by individual runs of the optimization algorithm with randomly initialized gates for PS circuits [cf.~\cref{psc_max_para}] with various $q$, for a target MPS [cf.~\cref{cls_mps}] with $\xi \approx 3.8$. The line connects the minimal $\kappa$ values achieved for each $q$, representing the exponential decay of the global minima $\kappa(q)$ with $q$, as observed in \cref{exp_pow_fig}(a). The inset shows the success probability $p_{\rm succ}$ of the optimization algorithm finding the global minimum $\kappa(q)$ within a relative error of $1\%$. We sample enough points to ensure the optimization meets the success criterion at least $10^3$ times for each $q$.}
        \label{scale_supp}
\end{figure}

To investigate the impact of local minima, in \cref{scale_supp}(b), we present the error density \( \kappa(q) \) obtained from individual runs of the optimization algorithm with randomly initialized gates for single-layer PS circuits with various \( q \) and \( l = q+1 \) [cf.~\cref{psc_max_para}]. As \( q \) increases (corresponding to a larger optimization problem size), we observe multiple instances where the optimization halts at local minima. We estimate the success probability \( p_{\rm succ}(q) \) of finding the optimal \( \kappa \) by calculating the ratio of optimizations that find the global minimum \( \kappa(q) \) within a relative error of \( 1\% \) and show the results in the inset of \cref{scale_supp}(b). Our results indicate that beyond a threshold value of \( q \), \( p_{\rm succ} \) decreases almost exponentially with \( q \). This behavior reflects the NP-hardness~\cite{bittel2021training} of classically optimizing variational circuits.

Remarkably, despite the near-exponential decay of \( p_{\rm succ}(q) \) with \( q \), our optimization algorithm is expected to run \( \text{poly}(N / \epsilon) \) times to find the global minimum, as \( q \) scales only logarithmically with \( N / \epsilon \). This suggests that the local optimization algorithm [cf.~\cref{algo_mps}] can scalably find sufficiently good solutions.

\subsection{Circuit compilation overhead of RG circuits~\cite{malz2023preparation} for preparing MPS of high bond dimensions}
\label{ps_rg_compare}

As discussed in the main text, by varying the number of layers \( M \), PS circuits can represent quantum states with entanglement across non-overlapping cuts bounded by \( S \leq 2M \). This has been explored in the case of multi-layer sequential circuits for preparing MPS with higher bond dimensions, such as the ground states of local Hamiltonians and MPS arising from algorithmic applications or quantum dynamics~\cite{Ran2020a,lin2021real,rudolph2023decomposition,iqbal2022preentangling,gundlapalli2022deterministic,melnikov2023quantum,iaconis2024quantum,bohun2024scalable,gonzalez2024efficient,sano2024quantum}. These studies generally suggest that a favorable number of layers, \( M = O(1) \), is sufficient to prepare relevant MPS with entanglement \( S = O(1) \) (where the bond dimension \( D \sim e^S \)). While the circuit depth for sequential circuits scales linearly with the system size as \( T_{\rm SEQ} \sim N \), PS circuits are expected to achieve comparable performance to sequential circuits with the same number of layers while requiring significantly shallower depths, as demonstrated for bulk-TI MPS with $D = 2$ and the ground state of the XY model [cf.~\cref{exp_pow_fig,xy_err_fig}].

On the other hand, there exists a rigorous algorithm to convert arbitrary MPS with high bond dimensions into a single dense ``layer'' of a sequential circuit~\cite{Schon2005} or to convert short-range correlated MPS into log-depth RG circuits~\cite{malz2023preparation}. It is already established that multi-layer sequential circuits (corresponding to PS circuits with $l = N - 1$ and $q = 1$) typically offer a more compact representation of high bond-dimension MPS compared to the `dense' sequential circuits~\cite{Ran2020a,rudolph2023decomposition,haghshenas2022variational,lin2021real}. To complement these results, we focus on analyzing the gate compilation overhead of RG circuits for preparing short-range correlated MPS with higher bond dimensions and qualitatively discuss why this makes PS circuits a preferable choice over RG circuits for preparing states with higher entanglement.

\subsubsection{$T_{\rm CNOT}$ scaling of the RG circuits}
\label{rg_review}

Here, we briefly review the RG circuits introduced in Ref.~\cite{malz2023preparation} and derive the scaling of \( T_{\rm CNOT} \) for preparing an \( N \)-qubit (\( d = 2 \)) MPS [cf.~\cref{mps_form}] with bond dimension \( D \) and correlation length \( \xi \), up to a small infidelity \( \epsilon \). Note that the analysis below can be easily generalized to physical dimension \( d \).

Starting from a given MPS, the RG circuits are constructed by grouping each set of \( q_b \) neighboring tensors \( \{A\} \) into a single tensor \( B \), followed by a polar decomposition \( B = V_{\rm RG}P_{\rm RG} \) to extract the isometry \( V_{\rm RG} \), which renormalizes the MPS toward its fixed point~\cite{malz2023preparation}.
\begin{align}
\begin{array}{c}
  \begin{tikzpicture}[scale=.45,thick,baseline={([yshift=-3ex]current bounding box.center)}]
    \BTensor{0,0}{$B$}
  \end{tikzpicture}
\end{array}
&=
\begin{array}{c}
  \begin{tikzpicture}[scale=.45, baseline={([yshift=5.5ex]current bounding box.center)}, thick]
    \draw[shift={(0,0)},dotted] (0,0) -- (4,0);
    \ATensor{0,0}{$A$}
    \ATensor{4,0}{$A$}
    \draw [decorate, decoration = {calligraphic brace,mirror}] (0,-0.8) --  (4,-0.8);
    \draw (2,-1.5) node {\scriptsize $q_b$};
  \end{tikzpicture}
\end{array}
\nonumber \\
\begin{array}{c}
  \begin{tikzpicture}[scale=.5,thick]
    \BTensor{0,0}{$B$}
  \end{tikzpicture}
\end{array}
&=
\begin{array}{c}
  \begin{tikzpicture}[scale=.4,thick]
    \PTensor{0,0}
    \isometry{0,1.5}{$V_{\rm RG}$}
  \end{tikzpicture}
\end{array}
\label{eq:rg_decomposition}
\end{align}
Therefore, one can construct an approximation of the target MPS as 
\begin{align}
|\widetilde \phi_N \rangle 
&= \left(\bigotimes_{i=1}^{N/q_b} U_i\right) 
   \bigotimes_{i = 1}^{N/q_b} 
   \left( \ket{\omega}_{R_i L_{i+1}} \ket{0 \dots 0}_{C_i} \right) \nonumber \\
&=
\begin{array}{c}
  \begin{tikzpicture}[scale=.5,thick]
    \foreach \x in {0,1,...,1}{
      \unitary{5*\x, 1}
      \draw (1.2+5*\x, 0) -- (5*\x + 3.8, 0);
      \filldraw[color=black, fill=white, thick](5*\x+2.5, 0) circle (0.3);
    }
    \draw (-1.2, 0) -- (-2.1, 0);
    \draw[dotted] (9,  0) -- (10, 0);
    \draw[dotted] (-2.2,  0) -- (-3.3, 0);
    \draw (-1.3,-1.2) node {\scriptsize $L_i$};
    \draw (0.05,-1.2) node {\scriptsize $C_i$};
    \draw (1.5,-1.2) node {\scriptsize $R_i$};
    \draw (3.7,-1.25) node {\scriptsize $L_{i+1}$};
    \draw (5.15,-1.25) node {\scriptsize $C_{i+1}$};
    \draw (6.75,-1.25) node {\scriptsize $R_{i+1}$};
  \end{tikzpicture}
\end{array}
\label{eq:phi_tilde}
\end{align}
where the unitary is constructed such that it implements the required isometry when acting on a product state  $\ket{0}^{\otimes (q_b-2)}$ over the `central' region ($q_b = 4$ in the illustration below), and the fixed-point state $| \Omega \rangle = \bigotimes_{i=1}^{N/q}  \ket{\omega^i}_{R_i L_{i+1}} $ is a tensor product of entangled pairs of qudit with qudit dimension $D$,
\begin{align}
\begin{array}{c}
  \begin{tikzpicture}[scale=.4,thick,baseline={([yshift=3.8ex]current bounding box.center)}]
    \unitary{0,0}
  \end{tikzpicture}
\end{array}
&=
\begin{array}{c}
  \begin{tikzpicture}[scale=.4,thick,baseline={([yshift=0ex]current bounding box.center)}]
    \isometry{0,0}{$V_{\rm RG}$}
  \end{tikzpicture}
\end{array},
\nonumber \\
|\omega\rangle_{R_i L_{i+1}} &=
\begin{array}{c}
  \begin{tikzpicture}[scale=.45,thick,baseline={([yshift=4ex]current bounding box.center)}]
    \draw (-1, .5) -- (-1, 0) -- (1,0) -- (1,0.5);
    \filldraw[color=black, fill=white, thick](0, 0) circle (0.3);
    \draw (-.8,-0.7) node {\scriptsize $R_i$};
    \draw (1.25,-0.75) node {\scriptsize $L_{i+1}$};
  \end{tikzpicture}
\end{array}
\label{eq:rg_pair}
\end{align}
The infidelity $\epsilon$ between the approximation $|\widetilde \phi_N \rangle$ and the target MPS is shown to scale as $\epsilon=O\left( \frac{N}{q} e^{-\gamma q/\xi} \right)$~\cite{Piroli2021a,malz2023preparation}, which is numerically demonstrated for bulk-TI MPS of $D=2$ in \cref{exp_pow_fig}(a)~\footnote{For a given $q$ of the PS circuit, we plot the $q_b = q + 1$ result of the RG circuit in \cref{exp_pow_fig}(a). This choice corresponds to comparing the RG circuit and the PS circuit with the same maximal correlation range [cf.~\cref{ps_cor_illu}].}

\begin{figure*}
	\centering
	\includegraphics[width=0.98\textwidth]{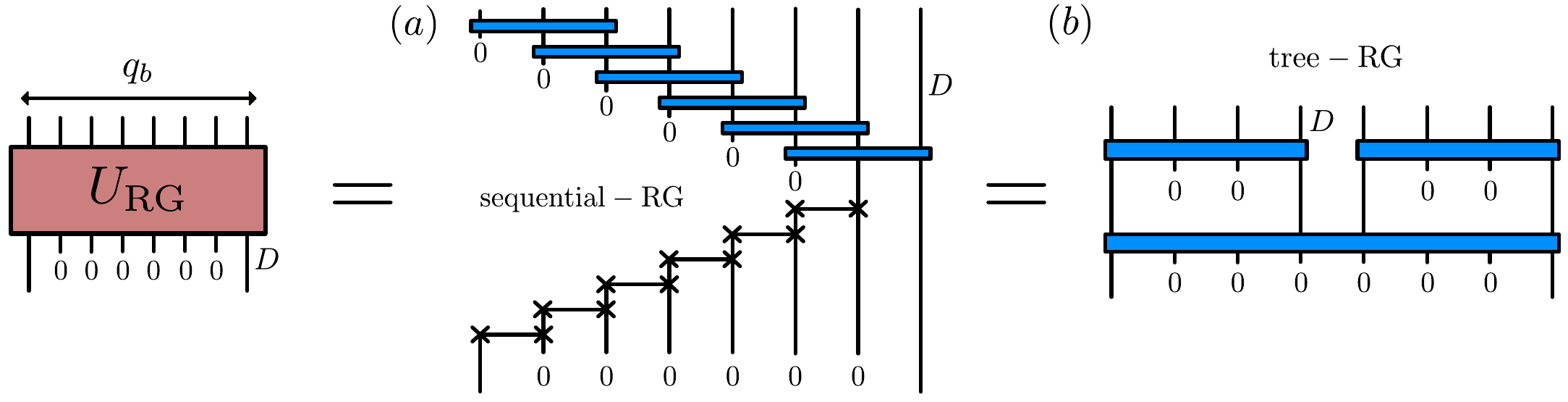}
        \caption{Decomposition of the isometry $U_{\rm RG}$ as (a) the sequential-RG circuit, or (b) tree-RG circuit. The `0' denotes the input product state $|0\rangle$ of the isometries. See Ref.~\cite{malz2023preparation} for details.}
        \label{rg_circ_fig}
\end{figure*}

Ref.~\cite{malz2023preparation} provides two explicit constructions for the circuits within the unitary $U_{\rm RG}$, illustrated in \cref{rg_circ_fig}. The \textit{sequential-RG} circuit iteratively applies singular value decompositions to construct a sequential circuit on $q_b$ sites, consisting of multi-qubit isometries with a mapping dimension of $D^2 \rightarrow 2D^2$ and a layer of $O(q)$ SWAP gates. In contrast, the \textit{tree-RG} approach iteratively applies blocking and polar decomposition for $O(\log \log N)$ times, yielding a tree-like circuit with $O(\log \log N)$ layers of long-range isometries, each with a mapping dimension of $D^2 \rightarrow D^4$. Taking $k = \lceil \log_2 D \rceil$,

By using the lower bound from \cref{iso_tcnot} for the CNOT depth required to implement an isometry, we can derive a lower bound of the total CNOT depth, $T_{\rm CNOT}$, for the RG circuit with a blocking size of $q_b>2k$ as:
\begin{equation} \label{}
	T_{\rm CNOT}^{\rm Seq-RG} \geq T_{\rm pair} + T_{\rm SWAP} + T_{\rm up},
\end{equation}
where each component is given by:
\begin{align}
T_{\rm pair} &= T_{\rm iso}(0, 2k) \nonumber \\
             &= \left\lceil \frac{1}{4} \left(2^{2k+1} - 4k - 2\right) \right\rceil \sim D^2
\end{align}
\begin{equation} \label{}
T_{\rm SWAP} = 3(q_b-k-1),
\end{equation}
\begin{align}
T_{\rm up} &= (q_b - 2k)\, T_{\rm iso}(2k, 2k+1) \nonumber \\
           &= (q_b - 2k)\left\lceil \frac{1}{4} \left(3 \cdot 2^{4k} - 6k - 3\right) \right\rceil \sim D^4
\end{align}
For the tree-RG circuits, the total CNOT depth is:
\begin{equation} \label{}
	T_{\rm CNOT}^{\rm tree-RG} \geq T_{\rm pair} + T_{\rm tree}
\end{equation}
\begin{align}
T_{\rm tree} &= (q_b - 2k)\, T_{\rm iso}(2k, 4k) \nonumber \\
             &= (q_b - 2k) \left\lceil \frac{1}{4} \left(2^{6k+1} - 2^{4k} - 10k - 1 \right) \right\rceil \sim D^6
\end{align}
Therefore, for $D\gg 1$, we asymptotically have $T_{\rm CNOT}^{\rm Seq-RG} = {\tilde \Omega}(D^4) $, and $T_{\rm CNOT}^{\rm tree-RG} ={\tilde \Omega} (D^6)$. Here the notation ${\tilde \Omega}$ suppresses the logarithmic factors. 

The above scalings suggest that RG circuits incur a significant circuit compilation overhead for preparing short-range correlated MPS with higher bond dimensions. As a simple example, achieving an MPS bond dimension of $D=64$ corresponds to an approximate $10^6$-fold (sequential-RG) and $10^9$-fold (tree-RG) increase in $T_{\rm CNOT}$ for fixed $q_b$, rendering RG circuits impractical for preparing MPS with high bond dimensions. In contrast, a PS circuit with $M=3$ layers can already prepare a certain subclass of MPS with a bond dimension of $D=2^{2M}=64$. Although it remains to be explored how the number of parameters scales for PS circuits to prepare generic MPS with high bond dimensions, we expect that PS circuits will be a better choice than RG circuits for preparing states with higher entanglement on near-term noisy quantum devices.
 
We also note that it is possible to consider circuit layouts formed by multiple layers of RG circuits for $D = 2$ to reduce the circuit depth required for RG circuits to create entanglement. However, since the single-layer PS circuits achieve the same performance as RG circuits while having a shallower depth [cf.~\cref{exp_pow_fig}(a,b)], we expect that multi-layer PS circuits will also outperform such multi-layer RG circuits for MPS preparation.

\section{Noise robustness of PS circuits}
In this section, we provide additional details on the noise robustness of PS circuits, including error propagation, gradient scaling, and evaluation accuracy.

\subsection{The impact of noise on the energy density}
\label{apd_noise_E}
Given a circuit layout, we use a DMRG-like algorithm~\cite{evenbly2009algorithms} to determine the circuit parameters ${\bm \theta}_{\rm best}$ that minimize the energy  
\begin{equation} \label{}
E = \langle \Psi_{\rm PS}({\bm \theta}_{\rm best}) | H_{XY} | \Psi_{\rm PS}({\bm \theta}_{\rm best}) \rangle	
\end{equation}
 for a system size of \( N = 30 \), ensuring that finite-size effects are negligible. We compare the ground-state energy error density, \( \nu_{\rm XY} \equiv (E - E_{\rm GS}) / N \), where \( E_{\rm GS} \) is the exact ground-state energy, with the results shown in \cref{xy_err_fig}(a). 

\begin{figure*}
	\centering
	\includegraphics[width=0.88\textwidth]{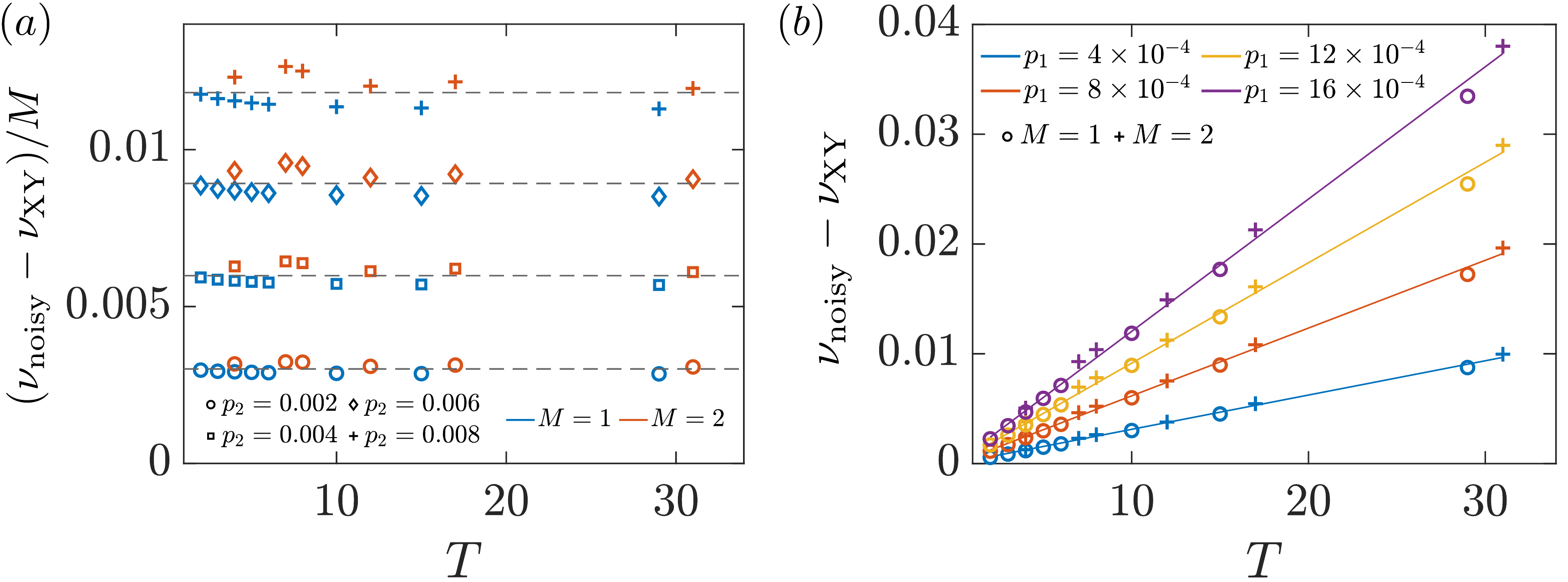}
        \caption{Numerical verification of the scaling of the energy density difference between noisy and noiseless PS circuits, \( \nu_{\rm noisy} - \nu_{\rm XY} \) [\cref{err_scale}], for the cases of (a) various circuit layouts with two-qubit gate errors (\( p_1 = 0 \)) and (b) various circuit layouts with idling errors (\( p_2 = 0 \)).}
        \label{energy_verify}
\end{figure*}

To evaluate the impact of noise, we introduce errors following a noise model that includes both idling errors and two-qubit gate errors [cf.~\cref{xy_err_fig}(b)]. We then calculate the energy error density 
$ \nu_{\rm noisy} \equiv (E_{\rm noisy} - E_{\rm GS})/{N}$, where $ E_{\rm noisy} = {\rm Tr} [\rho({\bm \theta}_{\rm best}) H_{XY}]$, and \( \rho({\bm \theta}_{\rm best}) \) is the density matrix generated by the noisy circuit using the same optimized parameters \( {\bm \theta}_{\rm best} \). The results are shown in \cref{energy_verify}, confirming the linear dependence of the energy density on the amount of noise present in the circuits, as [cf.~\cref{err_scale}, restated here]:
\begin{equation} \label{err_scale}
\nu_{\rm noisy} - \nu_{\rm XY} \approx  c_{E} (p_1 T + 2 p_2 M).
\end{equation}
where $c_{E}$ is a model-dependent constant ($c_{E} \approx 0.7$ for the XY model).

\subsection{Details of Gradiant variance calculation}
\label{apd_vqe_form}

\begin{figure}[h!]
	\centering
	\includegraphics[width=0.48\textwidth]{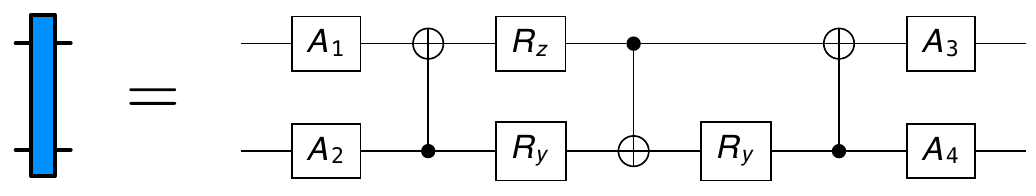}
        \caption{Circuit decomposition of an arbitrary two-qubit gate. Here $A_1, \ldots, A_4$ are arbitrary single-qubit rotations, and $R_z, R_y$ are single-qubit rotations along the $z$ and $y$ directions [cf.~\cref{single_A}].}
        \label{twob_circ}
\end{figure}

To study the scaling of gradient variance for noisy PS circuits, we parameterize each two-qubit gate using CNOT gates and single-qubit rotations~\cite{Barenco1995}, as shown in \cref{twob_circ}. An arbitrary single-qubit unitary $A$ is parametrized as
\begin{align}
A(\bm \theta) &= e^{i \alpha} R_{z}(\beta)\, R_{y}(\gamma)\, R_{z}(\delta) \nonumber \\
\text{with} \quad R_{\alpha}(\theta) &= \exp(-i \sigma_\alpha \theta / 2)
\label{single_A}
\end{align}
where $\{\sigma_\alpha\}_{\alpha = x, y, z}$ are Pauli matrices. Thus, each two-qubit gate contains 15 parameters, and all parameters of the circuit are collectively denoted as $\bm \theta$. We use automatic differentiation implemented in PastaQ.jl~\cite{pastaq} to evaluate the gradient variance of the cost function $E_{\rm noisy}({\bm \theta}) = {\rm Tr} [\rho({\bm \theta}) H_{XY}]$ for individual random realizations of the circuit. The gradient variance is then averaged over all realizations and parameters $\bm \theta$ to compute 
\begin{equation} \label{}
V_E^{\rm noisy}(M, T, p_1, p_2) \equiv \langle (\partial_{\theta_j} E_{\rm noisy})^2 \rangle_{\bm \theta} - \langle \partial_{\theta_j} E_{\rm noisy} \rangle^2_{\bm \theta},	
\end{equation}
as numerically illustrated in \cref{err_fig}(a).

\begin{figure*}
	\centering
	\includegraphics[width=\textwidth]{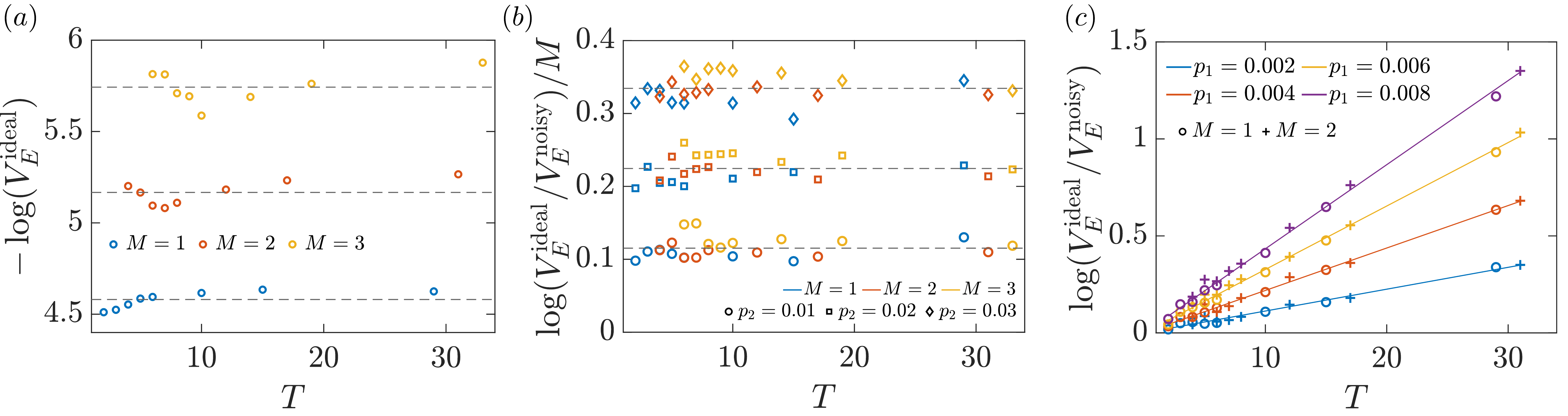}
        \caption{Numerical verification of the scaling of the gradient variance $V_E$ [cf.~\cref{ve_scale}]. (a) The scaling of the gradient variance in noiseless circuits (denoted as $V_E^{\rm ideal}$) for various circuit depths $T$ and numbers of layers $M$. The dashed lines represent the mean value of the data for the same $M$. (b) The scaling of the ratio between noisy $V_E$ and noiseless $V_E^{\rm ideal}$ for various $T$ and $M$, with different two-qubit gate noise levels $p_2$, and fixed $p_1 = 0$. We also perform a data collapse by dividing the exponent by $M$. The dashed lines represent the mean values of the data for the same $p_2$. (c) Same as (b), but for various $p_1$ and fixed $p_2 = 0$. The lines represent linear fits to the data for the same $p_1$.}
        \label{grad_verify}
\end{figure*}

In \cref{grad_verify}(a), we verify the scaling of $V_E^{\rm ideal}$ for noiseless PS circuits with $q = 1$ [cf.~\cref{ep_apd_fig}(d)]~\footnotemark[3], and \cref{grad_verify}(b, c) confirms its scaling with the amount of error density $p_1 T$ and $2 p_2 M$. The prefactors for the XY model are $\alpha \approx 0.6$ and $c_V \approx 5.5$.  Therefore, \cref{ve_scale} provides a good qualitative description of the gradient variance in PS circuits and suggests that PS circuits with constant $M$ layers exhibit better trainability than sequential circuits and brickwall circuits of the same depth, as discussed in the main text [cf.~\cref{err_fig}(a)].

\subsection{Details of error propagation calculation}
\label{apd_ep_calc}

\subsubsection{Review on the error propagation setup in Ref.~\cite{PRXQuantum.3.040326}}

As mentioned in the main text, to study the effect of error propagation, we consider the setup proposed in Ref.~\cite{PRXQuantum.3.040326}, which we briefly review here, and refer the reader to Ref.~\cite{PRXQuantum.3.040326} for more details.

We consider a noisy random circuit evolution that begins from an initial product state $|0\rangle^{\otimes N}$, followed by an entangling evolution $U$ and the corresponding disentangling evolution $U^\dagger$, while experiencing single-qubit depolarizing noise with a rate of $p_1$. Here, $U$ represents a random circuit of depth $T$ with a given layout; the circuit depth of $U^\dagger U$ is consequentially $2T$.

In the absence of errors, the final state is again $\ket0^{\otimes N}$. If a single-qubit depolarizing error occurs somewhere in the circuit, it may end up affecting more than a single qubit, because the random gates can propagate errors.
In particular, if we consider the average channel obtained by acting with a two-qubit gate $\mathcal U$ some channel $\mathcal M$ and the inverse gate $\mathcal U^\dagger$
\begin{equation} \label{}
	\mathcal{E}_d(\rho)=\int_{\mathcal{U}} d \mathcal{U}\left[\mathcal{U}^{\dagger} \mathcal{M} \mathcal{U}\right](\rho),
\end{equation}
we find that depending on whether $\mathcal M$ depolarizes zero, one, or two qubits, there are the three possible outcomes~\cite{PRXQuantum.3.040326}:
\begin{enumerate}
	\item If no error occurs, i.e. $\cal M(\rho) = \rho$, then $\mathcal{E}_d(\rho) = \rho$.
	\item If both qubits are depolarized, i.e., $\mathcal{M}(\rho)=\operatorname{Tr}(\rho) \mathbbm{1}^{\otimes 2} / 4$, then $\mathcal{E}_d(\rho)=\operatorname{Tr}(\rho) \mathbbm{1}^{\otimes 2} / 4$.
	\item If an error occurs in only one of the qubits, i.e., $\mathcal{M}(\rho)=\mathbbm{1} / 2 \otimes \operatorname{Tr}_1(\rho)$ or $\mathcal{M}(\rho)= \operatorname{Tr}_2(\rho) \otimes \mathbbm{1} / 2 $, then 
\begin{equation} \label{ep_prob}
\mathcal{E}_d(\rho)=\frac{1}{5} \rho+\frac{4}{5} \operatorname{Tr}(\rho) \frac{\mathbbm{1}^{\otimes 2}}{4}.	\nonumber
\end{equation}
\end{enumerate}
Equation \ref{ep_prob} describes the error propagation, where there is a probability of $4/5$ that a single error evolves into two errors after being processed by the random unitary and a $1/5$ chance that it disappears. This enables modeling the error propagation process as a Markov chain, where the system is represented as a string of zeros and ones: zeros denote noiseless qubits, and ones denote depolarized qubits. Using Markov-chain Monte Carlo, the average number of depolarized qubits, $\langle \eta \rangle$, can be computed at the end of the circuit evolution~\cite{PRXQuantum.3.040326}, as shown in ~\cref{err_fig}(b).

In the following, we show that the ratio of depolarized qubits $\langle \eta \rangle /N$ in multi-layer sequential circuits is dominated by a term proportional to $p_1MT$ for $M>1$, which notably coincides with the behavior of brickwall circuits, where $M=T/2$ and $\langle \eta \rangle /N \propto p_1 T^2$.
We then confirm numerically that PS circuit exhibit the same behavior. Thus for $p_1MT\ll1$, we find that $\langle \eta \rangle$ interpolates between the brickwall and sequential limits and find that it is well approximated as
\begin{equation}\label{ep_scale_apd}
    \langle\eta\rangle/N \approx p_1 T (c_{\eta,1} + c_{\eta,2} M),
\end{equation}
with prefactors $c_{\eta,1},c_{\eta,2}$ of order 1 and mildly dependent on $T, M$, but very weak dependence on system size.

\subsubsection{Verification of the error propagation scaling [\cref{ep_eq}]}

\begin{figure*}
	\centering
	\includegraphics[width=0.98\textwidth]{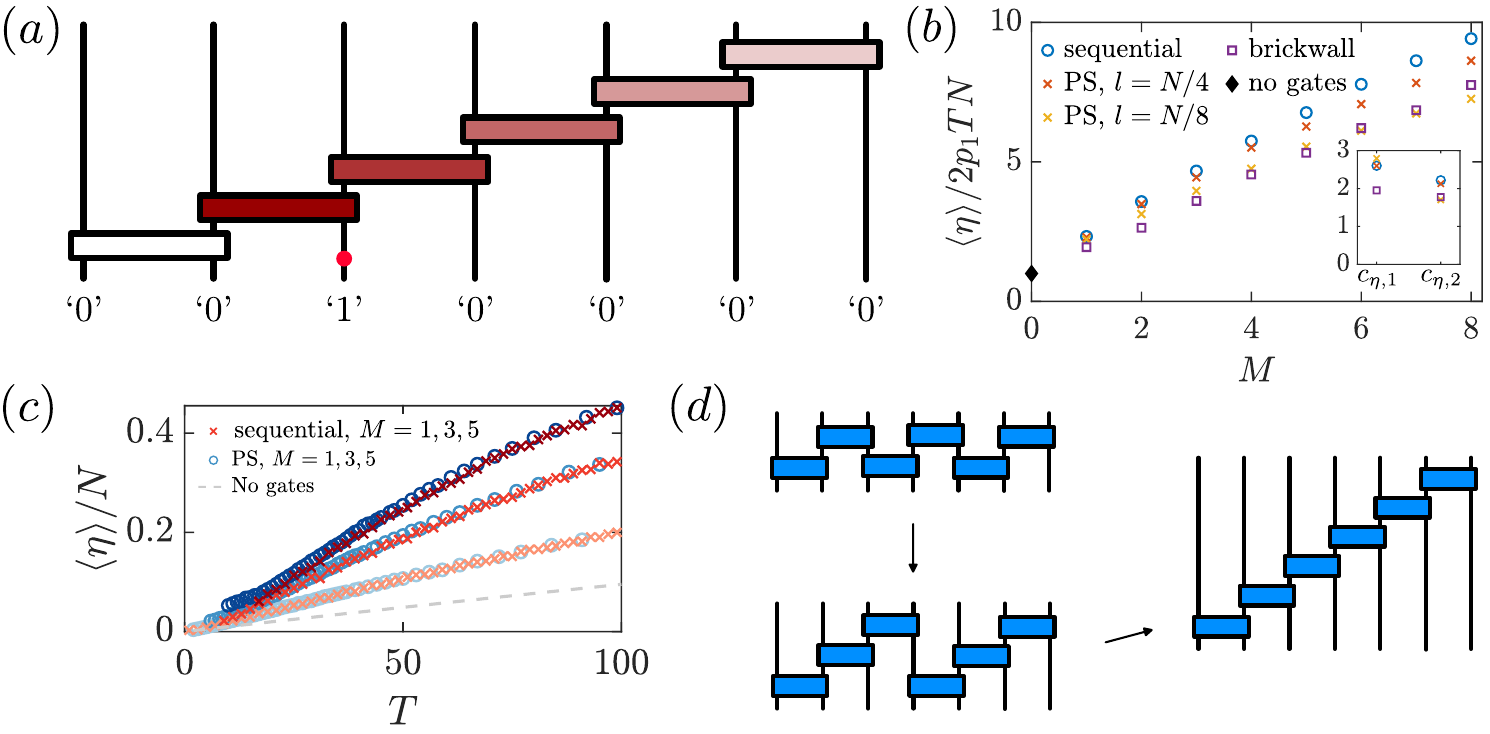}
        \caption{(a) Illustration of the propagation of a single error (denoted as `1' and the red dot) in the single-layer sequential circuit. The brown color represents the gates that propagate errors following the Markov-chain evolution~\cref{ep_prob}, with gradually lighter shades indicating that the probability of error propagation decreases with distance. (b) Scaling of the ratio of depolarized qubits, $\langle \eta \rangle / N$ (normalized by $2p_1T$), with the number of layers $M$ for PS circuits with different layouts, including the sequential and brickwall circuits. The inset show the extracted coefficient $c_{\eta,1}$ and $c_{\eta,2}$ [cf.~\cref{ep_scale_apd}].  (c) Scaling of $\langle \eta \rangle / N$ with circuit depth $T$ for PS circuits [same data as in ~\cref{err_fig}(b)] and sequential circuits, where $T$ is scaled by varying the system size $N$, with $T = N + 2M - 3$. The noise rate $p_1 = 5 \times 10^{-4}$. (d) Illustration of PS circuit layouts ($q = 1$) interpolating between brickwall and sequential circuits by tuning the chunk length $l$.}
        \label{ep_apd_fig}
\end{figure*}

\textbf{Derivation and evidence for the sequential circuits.---}We first consider a single depolarizing error that occurs in the single-layer sequential circuit at site $i$ in bulk, as illustrated in \cref{ep_apd_fig}(a). Following the Markov chain evolution described in \cref{ep_prob}, each two-qubit gate receiving one depolarized qubit as input propagates the error with a probability of $4/5$ (successful), while the error disappears with a probability of $1/5$ (unsuccessful). From the layout of the circuit [cf.~\cref{ep_apd_fig}(a)], it is evident that error propagation leads to a string of errors whose length depends on how often the error was propagated successfully. 

Given a error string of arbitrary length $k$, applying a sequential circuit extends this string by an amount $r\in\{-1,0,1,2,\dots\}$ with probability $(4/5)^{r+1}/5$. On average, the length of the string thus increases by
\begin{equation}
	\frac{1}{5} \sum_{k=0}^\infty \left(\frac{4}{5}\right)^k \cdot (k-1) = 3
\end{equation}
for each application of a sequential circuit.
However, if the string length hits 0, the dynamics gets stuck there, since subsequent sequential circuits do nothing to the string length. This dynamics is similar to a biased random walk with absorbing boundary condition at 0.
In our case, the probability of being absorbed at 0 approaches 1/4 as $M$ approaches infinity.
After the $M^{\mathrm{th}}$ sequential circuit, we thus have a probability distribution of error string lengths $l$ and that yields some expected error string length $L(M)$.
The first few values of $L(M)$, are $L(0)=1$, $L(1)=4$, $L(2)=6.4$, $L(3)=8.704$,
$L(4)=10.97728$.

Asymptotically, we could estimate that the asymptotic error length is the probability of there still being an error ($3/4$) times the average length if there were no absorbing boundary: $L(M)\to(3/4)(1+3M)$. This result is almost correct, but it neglects the fact that the surviving trajectories grow faster than 3 at early times, because of the postselection on surviving. The true result to 10 digits of numerical precision in simulations for $M=50$ and $N=5000$ is
$L(M)\to 2+ (9/4)M$.

In the setup described in the main text, the system experiences idle errors at a rate of $p_1$~\footnote{Note that in this error propagation analysis, we set the two-qubit gate error $p_2 = 0$ to ensure a fair comparison between different circuit layouts, as we focus solely on the effects of error propagation here.}. For simplicity, we neglect the interaction between error propagation at different sites, which is valid in the regime of sparse errors $p_1 T M\ll 1$.
We also take $M\ll T$, which allows us to neglect errors produced a qubit between applying the first and last gate.
The qubit at location $j \in [2, N]$ accumulates a depolarizing probability $p_{d}(j) = 1-(1-p_1)^{2(j-1)}\approx 2p_1(j-1)$ before being acted upon by the gates in the sequential circuit [cf.~\cref{ep_apd_fig}(a)] ($p_d(1)=p_d(2)$, because the qubits are acted on simultaneously).
Thus, the average probability that a qubit has an error when it is hit by the first gate is $\approx p_1 T$.
Together with our result in the previous paragraph, the number of errors after $M$ layers of the sequential circuit will be $\approx p_1 T L(M)$. After the last gate of the $M$ layers was applied, the qubits idle for at time $N-j$, and during this time they may again depolarize, which happens with an average probability $p_1 T$. 

Putting everything together, we find that $\langle \eta \rangle/N \approx p_1 T(L(M)+1)$ for multi-layer sequential circuits.

\textbf{Extending to PS circuits.---}In ~\cref{err_fig}(b), we studied the error propagation effect in PS circuits with a constant number of layers $M$, where the depth $T$ is controlled by the chunk length $l$, with a fixed $q = 1$ (no extended gates)~\footnote{Since $M$-layer PS circuits with a finite $1 \leq q \leq l$ always contain fewer gates than the corresponding $M+1$-layer PS circuits with $q=1$, the results for $q=1$ are sufficient to represent the random circuit properties for other $q$.}. This setup corresponds to an interpolation between brickwall and sequential circuits, as illustrated in \cref{ep_apd_fig}(d). Notably, the PS circuit can be viewed as multiple parallelized sections of the sequential circuit covering the entire chain, where each section has a size of $l$. If $l$ is large, we expect the behavior to follow the one of sequential circuits, and when $l$ is small, it approaches the brickwall limit.
The dominant contribution to either limit is proportional to $p_1 MT$, with slightly different prefactors.
Ultimately, the behavior in \cref{ep_scale_apd} can serve as a good guide for general PS circuits, which we numerically verify in \cref{ep_apd_fig}(b, c).

\printbibliography

\end{document}